\documentclass[twocolumn,english,showpacs,preprintnumbers,amsmath,amssymb,floatfix]{revtex4-1}

\usepackage[T1]{fontenc}
\usepackage[latin9]{inputenc}
\usepackage{color}
\usepackage{array}
\usepackage{amstext}
\usepackage{graphicx}
\usepackage{esint}
\usepackage{rotating}
\usepackage[font=small,labelfont=bf]{caption}
\usepackage[colorlinks = true,
linkcolor = blue,
urlcolor  = blue,
citecolor = red,
anchorcolor = blue]{hyperref}

\makeatletter

\providecommand{\tabularnewline}{\\}

\@ifundefined{textcolor}{}
{%
 \definecolor{BLACK}{gray}{0}
 \definecolor{WHITE}{gray}{1}
 \definecolor{RED}{rgb}{1,0,0}
 \definecolor{GREEN}{rgb}{0,1,0}
 \definecolor{BLUE}{rgb}{0,0,1}
 \definecolor{CYAN}{cmyk}{1,0,0,0}
 \definecolor{MAGENTA}{cmyk}{0,1,0,0}
 \definecolor{YELLOW}{cmyk}{0,0,1,0}
 }

\@ifundefined{definecolor}
 {\usepackage{color}}{}
\@ifundefined{definecolor}
 {\usepackage{color}}{}
\makeatother
\usepackage{babel}

\begin{document}


\title { Determination of neutron fracture functions from a global QCD analysis of the leading neutron production at HERA  }

\author{ Samira Shoeibi$^{1}$ }
\email{Samira.Shoeibimohsenabadi@mail.um.ac.ir}

\author { Hamzeh Khanpour$^{2,3}$ }
\email{Hamzeh.Khanpour@mail.ipm.ir}

\author { F. Taghavi-Shahri$^{1}$ }
\email{Taghavishahri@um.ac.ir}

\author { Kurosh Javidan$^{1}$ }
\email{Javidan@um.ac.ir}

\affiliation {
$^{(1)}$Department of Physics, Ferdowsi University of Mashhad, P.O.Box 1436, Mashhad, Iran          \\
$^{(2)}$Department of Physics, University of Science and Technology of Mazandaran, P.O.Box 48518-78195, Behshahr, Iran         \\
$^{(3)}$School of Particles and Accelerators, Institute for Research in Fundamental Sciences (IPM), P.O.Box 19395-5531, Tehran, Iran    }

\date{\today}

%
%
\begin{abstract}\label{abstract}

In this article, we present our global QCD analysis of leading neutron production in deep inelastic scattering at H1 and ZEUS collaborations.
The analysis is performed in the framework of a perturbative QCD description for semi-inclusive processes which is based on the fracture functions approach.
Modeling the non-perturbative part of the fragmentation process at the input scale Q$_0^2$, we analyze the Q$^2$-dependence of the leading neutron structure functions and obtain the neutron fracture functions (neutron FFs) from next-to-leading order (NLO) global QCD fit to data.
We have also performed a careful estimation of the uncertainties using the ``Hessian method'' for the neutron FFs and corresponding observables originating from experimental errors.
The predictions based on the obtained neutron FFs are in good agreement with all data analyzed, at small and large longitudinal momentum fraction $x_L$ as well as the scaled fractional momentum variable $\beta$. 

\end{abstract}


\maketitle

\tableofcontents{}

%
\section{Introduction}\label{sec:Introduction}

Over the past decade, our knowledge of the quark and gluon substructure of the nucleon has been extensively improved due to the high-energy scattering data from fixed target experiments, the precise data from electron-proton collider HERA~\cite{Abramowicz:2016ztw,Abt:2016zth,Abramowicz:2015mha,Abramowicz:2014jak,Andreev:2013vha,Aaron:2009aa,Aaron:2009kv,Aaron:2009bp}, the data from high energy proton-proton scattering at the Tevatron~\cite{TevatronElectroweakWorkingGroup:2016lid,Abazov:2016kfr,Abazov:2016tba,Aaltonen:2016pos,D0:2016ull,Aaltonen:2008eq,Abulencia:2005yg,Abazov:2008ae,Abulencia:2007ez,Abbott:2000ew} and up-to-date data from LHC~\cite{CMS:2017tvp,Sirunyan:2017azo,Aad:2013lpa,Aad:2011fc,Chatrchyan:2013tia,Chatrchyan:2011wt,Aad:2013iua,Aaij:2012mda,Aaij:2015vua,Chatrchyan:2011jz,Aad:2011dm,Chatrchyan:2012xt}. Deep inelastic scattering (DIS) data as well as data from hadron colliders has been successfully used in many Global QCD analyses to extract the unpolarized parton distribution functions (PDFs)~\cite{Bourrely:2015kla,Ball:2012cx,Harland-Lang:2014zoa,Ball:2014uwa,Martin:2009iq,Gao:2013xoa,Alekhin:2012ig}, polarized PDFs~\cite{Shahri:2016uzl,Ball:2013lla,Jimenez-Delgado:2014xza,Sato:2016tuz,Nocera:2014uea,Leader:2014uua,Nocera:2014gqa}, nuclear PDFs~\cite{Khanpour:2016pph,Eskola:2016oht,Kovarik:2015cma,Klasen:2017kwb,Wang:2016mzo,deFlorian:2011fp}, and related studies~\cite{Goharipour:2017rjl,Carlson:2017gpk,Accardi:2011fa,Owens:2012bv,Monahan:2016bvm,Dahiya:2016wjf,Jimenez-Delgado:2013boa,Nocera:2016zyg,Ball:2016spl,Goharipour:2017uic,Ru:2016wfx,Haider:2016zrk,Accardi:2016qay,Armesto:2015lrg,Frankfurt:2015cwa,Guzey:2013xba,Frankfurt:2016qca,Guzey:2016qwo,Accardi:2011mz,Frankfurt:2011cs,Alekhin:2017kpj}.
Beside the mentioned data sets, the production of leading neutron and proton in deep inelastic scattering (DIS) opens a new window for the theory of strong interaction in the soft region and provides a probe of the relationship between QCD of quarks and gluons and the strong interaction of hadrons~\cite{Aaron:2010ab,Chekanov:2002pf}. Consequently, in the framework of perturbative QCD (pQCD), the study of leading-baryon production represents an important field of investigation. In the leading-baryon productions, $ep \to e^{\prime} B X$, the energetic neutron or proton  which are produced in the fragmentation of the proton remnant, carry a large fraction $x_L$ of the longitudinal momentum of the incoming proton
~\cite{Aaron:2010ab,Chekanov:2002pf,Adloff:1998yg,Breitweg:2000nk,Chekanov:2004wn,Aktas:2004gi,Chekanov:2004dk,Chekanov:2007tv,Chekanov:2008tn,Chekanov:2002yh,Andreev:2014zka}. These events are measured at small polar angle with respect to the collision axis.

However, due to difficulty of detecting the leading-baryon in high energy physics experiment, the data available are scarce. More recently the H1 and ZEUS collaborations at HERA have measured events, in which neutron is produced in the forward region, obtain sizable contributions of leading neutrons to the DIS cross sections, $\sim 8-10$\%~\cite{Aaron:2010ab,Chekanov:2002pf}. These kinds of processes open a new window to study hard processes in a new kinematical region to obtain information on soft quantum chromodynamics (QCD) dynamics. In that case, the measurements of leading-baryon structure functions can be used as a test of new aspects of QCD. Along with these experimental developments, the fracture functions approach has been developed in the framework of perturbative QCD in order to deal with such kind of forward processes~\cite{Trentadue:1993ka,Ceccopieri:2007th,Szczurek:1997cw}.

Fracture functions provides a QCD-based description of semi-inclusive DIS in the target fragmentation region. 
The formalism of fracture functions, where the leading particles production is described in terms of structure functions of the fragmented nucleon, has been successfully used to describe forward neutron data from the H1 and ZEUS collaborations~\cite{Ceccopieri:2014rpa,deFlorian:1998rj,deFlorian:1997wi}. As for DIS structure function, QCD can not predicted the shape of fracture functions.
As for parton distribution functions (PDFs), the non-perturbative neutron fracture function (neutron FFs) can be parameterized at a given initial scale Q$_0^2$. Fracture functions, the probabilities of finding a parton and a hadron in the target, can be related to the parton distributions of the object exchanged between the initial and final states. In the production of the leading neutron in the target fragmentation region this object is $\pi^+$ in the $e p \to e n X$ process. In the target fragmentation region, the corresponding cross sections is expressed as a convolution of the fracture functions, $M_{h/p}$, with the point like partonic cross sections.
In this paper, we present the results of our QCD global analysis of recent and up-to-date experimental data for the production of neutrons in the forward direction in DIS. As we mentioned, the results obtained in this analysis, is in the framework of fracture functions by modeling the neutron FFs at the input scale, Q$_0^2$. We propose a standard parametric form for the neutron FFs at a given initial scale Q$_0^2=1$ GeV$^2$ and obtain their parameters by next-to-leading order (NLO) global QCD fit to forward neutron production data measured by H1 and ZEUS collaboration at HERA. We find that our theory predictions are in satisfactory agreements with all data analyzed.

The outline of the present paper is the following. First, in Section~\ref{sec:Theoretical} we present the theoretical settings of the analysis. The details of the fitting methodology applied in this work and the functional forms used to extract neutron FFs are presented in Section~\ref{sec:analysis}. The details of the forward neutron production data from H1 and ZEUS collaboration are discussed in Section~\ref{sec:Data}.
Section~\ref{sec:uncertainties} provides the method of the $\chi^2$ minimization, uncertainties estimation and error calculations.
The results of present NLO neutron FFs fits and detailed comparison with available observables are discussed in Section~\ref{sec:Results}.
In Section~\ref{sec:LHC}, we briefly discuss the present and upcoming experimental data on the production of leading-baryons at LHC and at Jefferson Lab.
Finally Section~\ref{sec:Summary} contains the summary and conclusions.

%
\section{Theoretical framework}\label{sec:Theoretical}

We can now specify the theory settings used for the neutron FFs fits presented in this work. We will restrict ourselves to
a brief summary of the theoretical framework relevant for our global QCD analysis of leading neutron structure functions in which we closely follow Ref.~\cite{Ceccopieri:2014rpa}.

We use the NLO theory with $\alpha_s(M_Z^2) = 0.1184$ in variable flavour number scheme (VFNS) with charm and bottom masses of $m_c = 1.41$ and $m_b = 4.50$ GeV. 
In order to describe the hard scattering DIS process, we use the usual kinematic variables $x$, Q$^2$, $y$ in which are defined as
\begin{equation}\label{eq:kinematic}
	Q^2=-q^2, \, \, \, \, \, \, \, x=\frac{Q^2}{2p.q}, \, \, \, \, \, \, \, y=\frac{p.q}{p.k} \,,
\end{equation}
where in the DIS process, $p$ is the four-momenta of the incident proton, $k$ is the four-momenta of the incident positron and $q$ is the four-momenta of the virtual photon. 
The four-fold differential cross section to describe the baryon production processes $ep \to e^{\prime} B X$ can be obtained by semi-inclusive leading-baryon transverse and longitudinal structure functions, $F_2^{\rm LB(4)}$ and $F_L^{\rm LB(4)}$, which is defined as~\cite{Aaron:2010ab,Chekanov:2002pf}
\begin{eqnarray}\label{eq:cross-section1}
	\frac{d^4\sigma (ep \to e^{\prime} B X)}{d\beta \, dQ^2 \, dx_L \, dt} & = & \frac{4 \pi \alpha^2}{\beta Q^4} (1-y+\frac{y^2}{2}) F_2^{\rm LB(4)} (\beta, Q^2, x_L, t) \nonumber  \\
	& + &  F_L^{\rm LB(4)} (\beta, Q^2, x_L, t) \,.
\end{eqnarray}
The longitudinal momentum fraction $x_L$ and the scaled fractional momentum variable $\beta$ are defined by
\begin{equation}\label{eq:xL-beta}
	x_L \simeq \frac{E_B}{E_p}, \, \, \, \, \, \, \, \, \beta=\frac{x}{1-x_L} \,,
\end{equation}
where $x$ is the Bjorken variable, $E_p$ is the proton beam energy and $E_B$ is the energy of final-state baryon. In Eq.~\eqref{eq:cross-section1}, $t$ is the squared four-momentum transfer between the incident proton and the final state neutron. The $t$ integrated differential cross section can be obtained by
\begin{eqnarray}\label{eq:cross-section2}
	\frac{d^3\sigma (ep \to e^{\prime} B X)}{d\beta \, dQ^2 \, dx_L} & = & \int_{t_0}^{t_{min}} \frac{d^4\sigma (ep \to e^{\prime} B X)}{d\beta \, dQ^2 \, dx_L \, dt}  dt \nonumber  \\
	& = & \frac{4 \pi \alpha^2}{\beta Q^4} (1-y+\frac{y^2}{2}) F_2^{\rm LB(3)} (\beta, Q^2, x_L) \nonumber  \\
	& + & F_L^{\rm  LB(3)} (\beta, Q^2, x_L) \,,
\end{eqnarray}
where the integration limits are
\begin{eqnarray}\label{eq:integration-limits}
	t_{min} & = & -(1 - x_L) (\frac{m_N^2}{x_L} - m_p^2)\,, \nonumber  \\
	t_0 & = & t_{min} - \frac{(p_T^{max})^2}{x_L} \,.
\end{eqnarray}
$m_N$ is the mass of final-state baryon, $m_p$ is the proton mass, and $p_T^{max}$ is the upper limit of the neutron transverse momentum used for the $F_2^{\rm  LB(3)}$ measurement.
For the semi-inclusive processes which have final-state proton and neutron, the structure function $F_{2, L}^{\rm  LB(3)}$, is denoted by $F_{2, L}^{\rm  LP(3)}$ and
$F_{2, L}^{\rm LN(3)}$ respectively. In this paper which is correspond to a QCD analysis of forward neutron production, we define the reduced $e^+ p$ cross section $\sigma_r^{\rm  LN(3)}$ in term of leading neutron transverse $F_2^{\rm  LN(3)}$ and the longitudinal structure functions $F_L^{\rm  LN(3)}$ as~\cite{Aaron:2010ab,Chekanov:2002pf}
\begin{eqnarray}\label{eq:reduced}
	\sigma_r^{LN(3)} & = & F_2^{LN(3)} (\beta, Q^2, x_L)  \nonumber  \\ 
	& - & \frac{y^2}{1+(1-y)^2} F_L^{LN(3)} (\beta, Q^2, x_L)\,.
\end{eqnarray}

It is noteworthy to mention here that the leading neutron structure functions in above equations can be written in terms of neutron FFs and hard-scattering coefficients~\cite{Ceccopieri:2014rpa}. The Wilson coefficient functions are the same as in fully inclusive DIS~\cite{Vermaseren:2005qc}.

The well-known DGLAP evolution equations~\cite{Dokshitzer:1977sg,Gribov:1972ri,Lipatov:1974qm,Altarelli:1977zs} which are a set of an integro-differential equations can be used to evolve the polarized and unpolarized parton distributions functions to an arbitrary energy scale, Q$^2$.
The solutions of these evolution equations will provide us the valance, gluon, and sea quark distributions inside the nucleon.
These equations widely can be used as fundamental tools to extract the deep inelastic scattering (DIS) structure functions of proton, neutron and deuteron which enrich our current information about the structure of the hadrons.
Since the scale dependence of the cross section in forward particle production in DIS can be calculated within perturbative quantum chromodynamics (pQCD)~\cite{Trentadue:1993ka}, consequently the neutron fracture functions also obey the standard DGLAP evolution equations~\cite{Ceccopieri:2014rpa,Camici:1998bg}.

In Refs.~\cite{deFlorian:1998rj,Daleo:2003xg,Daleo:2003jf,Trentadue:1996jc,Camici:1998bg,Trentadue:1994iw} have been shown that, in the phenomenological level, the  
fracture functions well reproduce the leading proton data, thus one can use the common perturbative QCD approach to these particular classes of semi-inclusive processes.
So, like for the case of parton distributions functions (PDFs), one can use phenomenological model to describe forward neutron production and extract the neutron FFs from QCD fit to the data~\cite{Ceccopieri:2014rpa,Ceccopieri:2007th}.
The evolution equations of neutron FFs are easily obtained by DGLAP evolution equations~\cite{Trentadue:1993ka} as

\begin{eqnarray}\label{eq:DGLAP}
	Q^2 \frac{\partial M^B_{\Sigma/P} (\beta, Q^2, x_L)}{\partial Q^2}  & = & \frac{\alpha_s(Q^2)}{2 \pi}  \nonumber  \\  && \int_{\beta}^{1} \frac{du}{u} P_{\Sigma}^j(u) M^B_{\Sigma/P} (\frac{\beta }{u}, Q^2, x_L)\,, \nonumber  \\
	Q^2 \frac{\partial M^B_{g/P} (\beta, Q^2, x_L)}{\partial Q^2}  & = & \frac{\alpha_s(Q^2)}{2 \pi}   \nonumber  \\ && \int_{\beta}^{1} \frac{du}{u} P_{g}^j(u) M^B_{g/P} (\frac{\beta }{u}, Q^2, x_L)\,, \nonumber  \\
\end{eqnarray}

where $M^B_{\Sigma/P} (\beta, Q^2, x_L)$ and $M^B_{g/P} (\beta, Q^2, x_L)$ correspond to the singlet and gluon distributions, respectively~\cite{Ceccopieri:2007th}.
These non-perturbative distributions in which hereafter indicated by ``neutron FFs'' need to be parametrize at an input scale, Q$_0^2$. 
Their evolution to higher scale, $Q^2>Q_0^2$, can be described by using the evolution equation given above. 
$P_{\Sigma}$ and $P_{g}$ in Eq.~\eqref{eq:DGLAP} are the common NLO contributions to the splitting functions governing the evolution of unpolarized singlet and non-singlet combinations of quark densities in perturbative QCD. Splitting functions are perturbatively calculable as a power expansion in the strong coupling constant $\alpha_s$. The splitting functions $P_{\Sigma}$ and $P_{g}$ in Eq.~\eqref{eq:DGLAP} are the same as in fully inclusive DIS~\cite{Vogt:2004mw,Moch:2004pa,vanNeerven:2000uj,vanNeerven:1999ca,Moch:2014sna}

In the next sections, we give a detailed account of the first global analysis of neutron FFs performed in this study
which in the following will be referred to as ``{\tt SKTJ17}''. We first discuss in details the parameterization of neutron FFs and then we will present data selection and the determination of the best fit, which we compare to the fitted data. We then focus on the studies of uncertainties using the standard Hessian error matrix approach.

%
\section{NLO QCD analysis of neutron FFs and parameterization}\label{sec:analysis}

In order to obtain a parametrization for the neutron FFs, $\,\beta M^N_{i/P} (\beta, Q_0^2, x_L)$ with $i=\Sigma$ and $g$, at a given initial scale $Q_0^2$, we select a relatively simple functional dependence
in the variables $\beta$ and $x_L$ with enough flexibility as to reproduce the data accurately.
We assume the following general initial functional form at Q$_0^2$ = 1 GeV$^2$

\begin{eqnarray}\label{eq:PDFQ0}
	\beta M^N_{\Sigma/P} (\beta, Q_0^2, x_L) &=& {\cal A}_q(x_L) \, \beta^{a_q} (1 - \beta)^{b_q} ( 1 + c_q \, \beta )  \,, \nonumber \\
	\beta M^N_{g/P} (\beta, Q_0^2, x_L) &=& {\cal A}_g(x_L) \, \beta^{a_g} (1 - \beta)^{b_g} ( 1 + c_g \, \beta )  \,, \nonumber \\
\end{eqnarray}

where ${\cal A}_q(x_L)$ and ${\cal A}_g(x_L)$ define as

\begin{eqnarray}
	{\cal A}_q(x_L) = {\cal N}_q \, x_L^{A_q} (1 - x_L)^{B_q} ( 1 + C_q \, x_L^{D_q} )  \,, \nonumber \\
	{\cal A}_g(x_L) = {\cal N}_g \, x_L^{A_g} (1 - x_L)^{B_g} ( 1 + C_g \, x_L^{D_g} )  \,.
\end{eqnarray}

The label of $\,\Sigma/P$ and $\,g/P$ correspond to the singlet and gluon distributions, respectively. The $x_L$ dependence of the neutron FFs is encoded in ${\cal A}_i(x_L)$. Since the
present leading neutron data are not yet sufficient to distinguish $q (=u, d, s)$ from $\bar{q} (=\bar{u}, \bar{d}, \bar{s})$, we assume symmetric sea distributions throughout.
We will show that these kinds of parametrizations give relatively good initial approximations to the description of the H1 and ZEUS leading neutron
data sets~\cite{Aaron:2010ab,Chekanov:2002pf}, however, their survival seems unlikely in a more precise analysis. 
The available forward neutron production data are not accurate enough to determine all the shape parameters with sufficient accuracy. Eq.~\eqref{eq:PDFQ0} includes 18 free parameters in total in which we further reduce the number of free parameters in the final minimization.

The parameters $\{p_i\}$ representing our best global QCD fit of neutron FFs in Eq.~\eqref{eq:PDFQ0}, henceforth
denoted as {\tt SKTJ17} are given in Table~\ref{fit-parameters}.
A few additional remarks will be presented in Sec~\ref{sec:Results}. As we mentioned, the currently available leading neutron data do not fully constrain the entire $\beta$ and $x_L$ dependence of $\beta M^N_{\Sigma/P}$ and $\beta M^N_{g/P}$ imposed in Eq.~\eqref{eq:PDFQ0}. Consequently we are forced to make some restrictions on the parameter space $\{p_i\}$. We will return to this issue in a separate section.

Rather than determining also the strong coupling $\alpha_s(Q_0^2)$ in the global QCD fit along with the neutron FFs parameters, we fixed $\,\alpha_s(M_Z^2)$ value close to the updated Particle Data Group (PDG) average. The scale dependence of $\alpha_s$ is normally computed by numerically solving its renormalization group equation at next-to-leading order accuracy.
For the evolution we take $\,\alpha_s(M_Z^2) = 0.1184$~\cite{Olive:2016xmw,Agashe:2014kda}, and we choose to work in the variable flavor number scheme (VFNS) where charm and bottom quark distributions are radiatively generated from their corresponding thresholds~\cite{Martin:2009iq,Ball:2014uwa}. In the present analysis, all quarks are treated as massless and we fixed the heavy quark masses at $m_c = 1.41$ GeV, $m_b = 4.50$ GeV and $m_b = 175.0$ GeV. Our choice for the VFNS scheme is due to that for all presently available leading neutron observables, heavy quarks play a negligible role. 
The scale evolution equations for the neutron FFs are solved in $x$-space at next-to-leading order.
Likewise, all leading neutron observables used in our QCD fit are computed consistently at next-to-leading order accuracy in the $\overline{MS}$ factorization scheme.

%
\section{Leading neutron production data}\label{sec:Data}

Our first physics objective is to establish the set of neutron FFs that gives the optimum theoretical description of the available hard scattering leading neutron production data.
In this section, we will present the data sets used in the present analysis. 
The data sets that we will use is the following: The H1 data on the leading neutron production in DIS scattering~\cite{Aaron:2010ab} as well as the data from leading neutron production in $e^+p$ collision from ZEUS collaboration~\cite{Chekanov:2002pf}. The detail of the data sets will be presented in the next section.

\subsection{H1 data}\label{sec:H1Data}

\begin{figure}[htb]
	\begin{center}
		\vspace{0.5cm}
		\resizebox{0.450\textwidth}{!}{\includegraphics{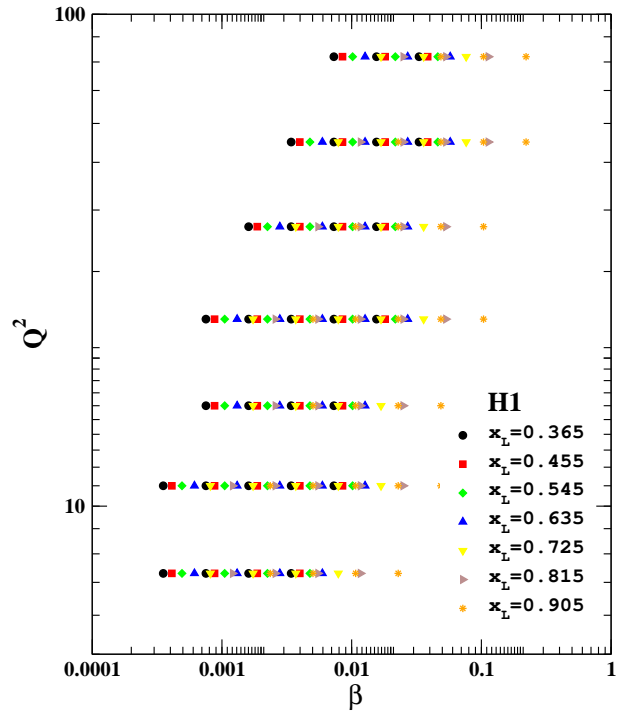}}   
		\caption{ (Color online) Nominal coverage of the H1 data sets used in our global QCD fits. The plot nicely summarizes the universal $\beta$, $x_L$ and $Q^2$ dependence of the forward neutron production at HERA~\cite{Aaron:2010ab}. For interpretation of the references to color in the figure legend, the reader is referred to the web version of
			this article.}\label{figH1-beta-Q-2}
	\end{center}
\end{figure}

The semi-inclusive cross sections data for the production of leading neutron are taken during the years of 2006 and 2007 by the H1 collaboration at HERA in DIS positron-proton scattering~\cite{Aaron:2010ab} which is correspond to an integrate luminosity of $\cal {L}$ = 122 pb$^{-1}$, much larger than the previous H1 measurement~\cite{Adloff:1998yg}. Better experimental capabilities in this measurement lead to the extension of the kinematical coverage of $x$ and $Q^2$ to higher values. 
This leading neutron structure function $F_2^{\rm LN(3)}$ in which has been measured by H1 experiment at HERA covers a large range of kinematics of $Q^2$, $\, 6 \leq Q^2 \leq 100$ GeV$^2$, and $x$, $\, 1.5 \times 10^{-4} \leq x \leq 3 \times 10^{-2}$, for average $y$ values between $0.05$ and $0.68$, and the upper limit of the neutron transverse momentum of $p_T^{\rm  max} < 200$ MeV. 
The value of longitudinal momentum fraction $x_L$ covers the range from $0.365$ to $0.905$. In order to enhance the relative contribution of pion exchange~\cite{Holtmann:1994rs,Kopeliovich:1996iw} for these selected of DIS events, the value of $p_T^{\rm max}$ in which used for the measurement of $F_2^{\rm LN(3)}$ is set to 200 MeV. Considering that the pion exchange mechanism dominates leading neutron production, these data sets can provide constraints on the shape of the pion structure function~\cite{McKenney:2015xis}.
In Fig~\ref{figH1-beta-Q-2}, we plot the nominal coverage of H1 data sets used in our QCD fits. The plot nicely summarizes the universal $\beta$, $x_L$ and $Q^2$ dependence of the forward neutron production at HERA. 

\subsection{ZEUS data}\label{sec:ZEUSData}

\begin{figure}[htb]
	\begin{center}
		\vspace{0.5cm}
		\resizebox{0.450\textwidth}{!}{\includegraphics{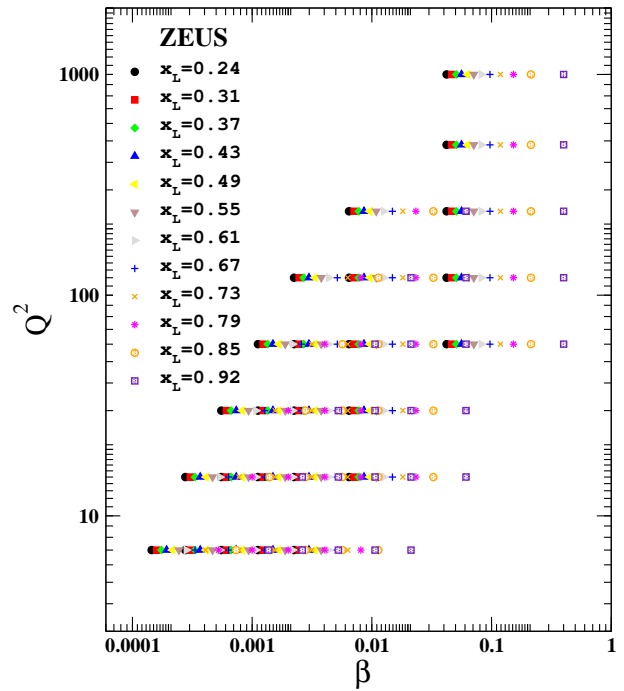}}   
		\caption{ (Color online) Nominal coverage of the ZEUS data sets used in our global fits~\cite{Chekanov:2002pf}. For interpretation of the references to color in the figure legend, the reader is referred to the web version of
			this article. }\label{figZEUS-beta-Q-2}
	\end{center}
\end{figure}

The semi inclusive cross section for production of leading neutron measured by ZEUS collaboration are also used in our global QCD analysis.
The ZEUS collaboration presented the leading neutron production cross sections for $x_L>0.2$ in neutral current electron-proton collisions at HERA~\cite{Chekanov:2002pf}. Positron and proton energies are $E_e= 27.5$ GeV and $E_p=820$ GeV, respectively, correspond to a center of mass energy of $\sqrt{s} = 300$ GeV. Similarly to H1 experiment, extensive range of kinematics was covered by the ZEUS data, for
$1.1 \times 10^{-4} \leq x \leq 3.2 \times 10^{-2}$ from photoproduction up to $Q^2 \sim 10^4$ GeV$^2$, with $0 < y < 0.8$ and neutron scattering angle $\theta_n < 0.8$ mrad. 
The HERA magnet apertures limit the FNC (Forward Neutron Calorimeter) acceptance to neutron with the production angle less than $\theta_n^{\rm max} = 0.8$ mrad, which is corresponding to the transverse momenta of $p_T^{\rm  max} = E_n \theta_n^{\rm max} = 0.656 x_L$ GeV.
As we already mentioned in the previous section, the distribution of the neutron for H1 data is integrated only up to $p_T^{\rm  max}= 200$ MeV, so the H1 and ZEUS data can only be used in the analysis for the longitudinal momentum fraction of $x_L= 0.3$, which is correspond to $p_T^{\rm  max} = 197$ MeV. 
For higher values of $x_L$, the ZEUS data should be scaled to account for the smaller $p_T$ range measured by H1 collaboration.
Of course, in general we would like to maximize the $\beta$, $x_L$ and $Q^2$ coverages included in the analysis in order to increase the statistics of our fit.
Therefore, we have scaled down the ZEUS data to the H1 $p_T$-range by using the form of $p_T^2$ distribution for the fixed values of $x_L$ as~\cite{Chekanov:2002pf}:

\begin{eqnarray}
 \frac{d \sigma^{\gamma^* p \to X n}}{d p_T^2} \propto e^{-b(x_L) \ p_T^2}\,,
\end{eqnarray}

where $\sigma^{\gamma^* p \to X n}$ is the the virtual photon-proton cross section for the process $\gamma^* p \to X n$.
The slope $b(x_L)$ can be parameterised as $b(x_L) = (16.3 \ x_L - 4.25)$ GeV$^{-2}$ which is in reasonable accord with the data~\cite{Chekanov:2002pf,Chekanov:2002yh}. In order to reduce the systematic uncertainties, ZEUS collaboration is measured the neutron-tagged cross section $ep \to e^{\prime} X n$ relative to the inclusive DIS cross section $ep \to e^{\prime} X$. Considering this ratio as well as proton structure function, one can obtain the $F_2^{\rm LN(3)}$ values for various bin of $x$, Q$^2$ and $y$. The kinematic range of ZEUS forward neutron data are shown in Fig.~\ref{figZEUS-beta-Q-2}. 
We should notice here that the H1 leading neutron data were collected during the 2006-2007 run by an integrated luminosity about 3 times that of the ZEUS data in the DIS region.
In consequence, the statistical uncertainties of the H1 data are much smaller than those for the ZEUS leading neutron spectra.

The measured leading neutron production data points above $Q^2$ = 1.0 GeV$^2$ used in the {\tt SKTJ17} global analysis are listed in Table~\ref{tab:tabdata}. 
For each data set we provide the corresponding references, the kinematical coverage of $x_L$, $x_B$ and $Q^2$, the number of data points and the fitted normalization shifts ${\cal{N}}_n$. 

%
\begin{table*}[htb]
	\caption{List of all the leading neutron production data points above $Q^2$ = 1.0 GeV$^2$ used in {\tt SKTJ17} global analysis. For each
		dataset we provide the corresponding references, the kinematical coverage of $x_L$, $x_B$ and $Q^2$, the number of data points and the fitted normalization shifts ${\cal{N}}_n$.} \label{tab:tabdata}
	\begin{tabular}{l c c c c c c}
		Experiments & [$x_L^{\rm min}, x_L^{\rm max}$] & [$x_B^{\rm min}, x_B^{\rm max}$]  & Q$^2$ GeV$^2$  & Number of data points & ${\cal N}_n$
		\tabularnewline
		\hline\hline
		H1~\cite{Aaron:2010ab} & [$0.365$--$0.905$]   & [$1.5 \times 10^{-4}$ -- $3.0 \times 10^{-2}$] & 7.3--82 & 203 & 0.9922  \\		
		ZEUS~\cite{Chekanov:2002pf}  &   [$0.240$--$0.920$] & [$1.1 \times 10^{-4}$ -- $3.2 \times 10^{-2}$] & 7--1000 & 300 &  1.0033 \\			
		\hline \hline
		\multicolumn{1}{c}{\textbf{Total data}}  &  &  & & \textbf{503} &    \\  \hline
	\end{tabular}
\end{table*}
%
%

%
\section{The method of $\chi^2$ minimization and neutron FFs uncertainties}\label{sec:uncertainties}

In this Section, we outline the details of {\tt SKTJ17} analysis. More specifically, we discuss the selection of data sets, treatment of experimental
normalization uncertainties, as well as the determination of the parameters by global $\chi^2$ minimization. 
We also briefly present the details of the Hessian matrix method for estimating uncertainties.
As we noted before, we have performed a careful estimation of the uncertainties using the ``Hessian method''.
An advantage of the Hessian technique is that it allows us to produce sets of eigenvector PDFs, which can be straightforwardly used in computations of other observables such as reduced $e^+ p$ cross section $\sigma_r^{\rm LN(3)}$ as well as leading neutron structure function $F_2^{\rm LN(3)}$.

\subsection{$\chi^2$ minimization}\label{sec:minimization}

Global QCD extractions of PDFs, nuclear PDFs as well as polarized PDFs are implemented around an effective $\chi^2$ function that quantifies the goodness of the fit to data for a given set
of theoretical parameters in which determines the PDFs at some input scale Q$_0^2$. In order to search for optimum PDFs by minimization, the simplest $\chi^2$ function is usually given by

\begin{equation}\label{eq:chi1}
\chi^2 (\{p_i\})  = \sum_{n=1}^{N^{\rm exp}} \sum_{j=1}^{N^{\rm data}_n} w_j \frac{~ ({\rm Data}_j - {\rm Theory}_j(\{p_i\})~)^2}{\delta \, {\rm Data}_j}   \,,
\end{equation}

The simple form of $\chi^2 (\{p_i\})$ presented above is appropriate only in the ideal case of data sets with uncorrelated errors. 
Since most experiments come with additional information on the fully correlated normalization uncertainty $\Delta {\cal N}_n$, Eq.~\eqref{eq:chi1} need to be modified in order to account for such normalization uncertainties.
In order to determine the best fit parameters of Eq.~\eqref{eq:PDFQ0}, we need to minimize the $\chi^2_{\rm global}$ function with the free unknown parameters.
$\chi_{\rm global}^2(\{p_i\})$ quantifies the goodness of fit to the data for a set of independent parameters $(\{p_i\})$ that specifies the neutron FFs at the input scale Q$_0^2$ = 1 GeV$^2$. This function is expressed as,
\begin{equation}\label{eq:chi2}
	\chi_{\rm global}^2 (\{p_i\}) = \sum_{n=1}^{N^{\rm exp}} w_n  \chi_n^2\,,
\end{equation}
where $w_n$ is a weight factor for the $n^{\rm th}$ experiment and
\begin{eqnarray}\label{eq:chi2global}
	\chi_n^2 (\{p_i\}) &=& \left( \frac{1 -{\cal N}_n }{\Delta{\cal N}_n}\right)^2  \nonumber \\
	&+& \sum_{j=1}^{N_n^{\rm data}} \left(\frac{ ( {\cal N}_n  \, {\rm Data}_j - {\rm Theory}_j(\{p_i\}) }{{\cal N}_n \, \delta {\rm Data}_j} \right)^2\,, \nonumber \\
\end{eqnarray}
where $N^{\rm exp}$ correspond to the individual experimental data sets and $N^{\rm data}_n$ correspond to the number of data points in each data set. 
The normalization factors $\Delta {\cal N}_n$ in Eq.~\eqref{eq:chi2global} can be fitted along with the fitted parameters $(\{p_i\})$.

The $\chi_{\mathrm{\rm global}}^2 (\rm p)$ function is minimized by the CERN program library MINUIT~\cite{James:1994vla}. From the $\chi_{\rm global}^2 (\{p_i\})$ analysis, an error matrix that is the inverse of a Hessian matrix is obtained. 
In order to determine the sensitivity of the fit to different values of $x_L$ collected by H1 and ZEUS collaborations at HERA, we compute the $\chi^2$ values for
each data sets.
The data sets included in {\tt SKTJ17} analysis are listed in Table~\ref{tab:chisquared}, together with the $\chi^2$ values, defined in Eq.\eqref{eq:chi1}, corresponding to each individual data set for each of $x_L$.  This suggests that reasonable fits to the leading neutron cross sections can be obtained within most of the $x_L$ values. More detailed discussion of the description of the individual data sets has been given in Section~\ref{sec:Data}.

%
\begin{table}[htbp]
	\centering
	{\footnotesize
		\begin{tabular}{c|c|c|c}
		\hline \hline
Experiment &   Data set    & $\chi^2$  & N$_{\rm pts}$    \\ \hline
		   & $x_L$ = 0.365 & 24.13       &    29           \\
   		   & $x_L$ = 0.455 & 25.62       &    29           \\
   		   & $x_L$ = 0.545 & 19.36       &    29           \\
H1 		   & $x_L$ = 0.635 & 19.28       &    29           \\
  		   & $x_L$ = 0.725 & 17.33       &    29           \\
   		   & $x_L$ = 0.815 & 13.23       &    29           \\
  		   & $x_L$ = 0.905 & 10.15       &    29           \\ \hline
All data sets & & \textbf{ 130.05 } & \textbf{ 203 }  \\    \hline
		   & $x_L$ = 0.240 & 24.84       &    25           \\
		   & $x_L$ = 0.310 & 9.68       &    25           \\
		   & $x_L$ = 0.370 & 11.68       &    25           \\
    	   & $x_L$ = 0.430 & 48.45       &    25           \\
   ZEUS	   & $x_L$ = 0.490 & 21.11        &    25            \\
		   & $x_L$ = 0.550 & 24.84       &    25           \\
		   & $x_L$ = 0.610 & 17.74       &    25           \\
		   & $x_L$ = 0.670 & 28.18       &    25           \\
		   & $x_L$ = 0.730 & 6.61       &    25           \\
		   & $x_L$ = 0.790 & 3.02       &    25           \\
		   & $x_L$ = 0.850 & 7.88       &    25           \\
		   & $x_L$ = 0.920 & 15.02       &    25           \\ \hline
All data sets & & \textbf{ 219.10 } & \textbf{ 300 }  \\
			\hline \hline
		\end{tabular}
	}
	\caption{The values of $\chi^2 / N_{\rm pts.}$ for the data sets included in the {\tt SKTJ17} global QCD analysis. More detailed discussion of the description of the individual data sets and the definitions of $\chi^2$ are contained in the text.}
	\label{tab:chisquared}
\end{table}

\subsection{Neutron FFs uncertainties}\label{sec:uncertainty}

As in the case of standard PDFs, the evolved leading neutron fracture functions are linear functions of the input densities.
Let  $M(\beta,Q^2,x_L;p_i|_{i=1}^k)$ be the evolved neutron FFs at $Q^2$ depending on
the parameters $p_i|_{i=1}^k$. Then its correlated error as given by Gaussian error propagation as~\cite{Bluemlein:2002be}
\begin{eqnarray}\label{uncertainties1} 
\Delta \beta M(\beta,Q^2,x_L) &=& \left\{ \sum_{i=1}^k \left( \frac{\partial \beta M}{\partial p_i} \right)^2 C(p_i,p_i)\right. \nonumber\\
&+&\left. \sum_{i\neq j=1}^k \left( \frac{\partial \beta M}{\partial p_i} \frac{\partial \beta M}{\partial p_j} \right) C(p_i,p_j) \right\}^{\frac{1}{2}}\,,  \nonumber\\
\end{eqnarray}
where $C(p_i,p_j)$ are the elements of the covariance matrix obtained in the QCD fit procedure at the input scale $Q_0^2$.
The covariance matrix can be used at any scale of $Q^2>Q_0^2$. The gradients $\partial \beta M/\partial p_i$ at this scale can be calculated
analytically. Their value at $Q^2$ is then calculated by evolution in $x$ space and are used according to Eq.~\eqref{uncertainties1}. 

In addition to the method presented above, one can also determine the uncertainties of obtained neutron FFs via well-known Hessian method and diagonalize the covariance matrix and work in terms of the eigenvectors and eigenvalues. Here, we briefly review the important points for studying the
neighborhood of $\chi^2_0$. The basic procedure is provided in Refs.~\cite{deFlorian:2011fp,Hou:2016sho,MoosaviNejad:2016ebo,Khanpour:2016uxh,Khanpour:2016pph,Martin:2002aw,Martin:2009iq,Pumplin:2001ct}.

As we have mentioned earlier, one can find the appropriate parameter set in which minimize the $\chi^2_{\rm global}$ function. We call this neutron FFs set
$S_0$. The parameters value of $S_0$, i.e. \{$p_1^0 \dots p_n^0$\}, in which extracted from QCD fit to H1 and ZEUS leading neutron data, will be presented in Sec.~\ref{sec:Results}. 
As we will mention latter, we simply fix some of the parameters of our input functional from presented in Eq.~\eqref{eq:PDFQ0} at their best-fit values, so that the Hessian matrix only depends on a subset of parameters.

By moving away the parameters from their obtained values, $\chi^2$ increases by the amount of $\Delta \chi^2$
\begin{eqnarray}\label{delta-chi} 
\Delta \chi^2_{\rm global} &=&  \chi^2_{\rm global}(\{p\}) - \chi^2_0(\{p^0\}) \nonumber \\
&& = \sum_{i, j=1}^n (p_i - p_i^0)  H_{ij}  (p_j - p_j^0)\, ,
\end{eqnarray}
where $H_{ij}$ is the Hessian matrix which defined as
\begin{equation}
H_{ij} = \frac{1}{2} \frac{\partial^2 \chi^2_{\rm global}} {\partial p_{i} \, \partial p_{j}} \Bigg|_{0}  \,.
\end{equation}
Now it is convenient to work in term of the eigenvalues and their corresponding orthogonal eigenvectors of covariance matrix. It is given by
\begin{equation}
\sum_{j=1}^n C_{ij} \upsilon_{jk} = \lambda_k \upsilon_{ik}\, ,
\end{equation}
and we should notice here that $C_{ij} \equiv H_{ij}^{-1}$ is the error (or covariance) matrix.
The displacement of the parameter \{$p_i$\} from its obtained minimum $p_i^0$ can be expressed in terms of the rescaled eigenvectors $e_{ik} = \sqrt {\lambda_k} \, v_{ik}$, that is
\begin{equation}\label{zk}
p_i - p_i^0 = \sum_{k=1}^n \, e_{ik} \, z_k \, .
\end{equation}
Considering the orthogonality of eigenvectors $\upsilon_{ik}$ and putting Eq.~\eqref{zk} in \eqref{delta-chi}, one can write
\begin{eqnarray}\label{delta-chi5} 
\Delta \chi^2_{\rm global} =  \chi^2_{\rm global}(\{p\}) - \chi^2_0(\{p^0\})
 = \sum_{k=1}^{n} \, z_k^2\, .
\end{eqnarray}
The relevant neighborhood of $\chi^2$ is the interior of hypersphere with radius $T$. This means that
\begin{equation}\label{tt}
\sum_{k=1}^nz_k^2 \leq T^2\, .
\end{equation}
Finally the neighborhood parameters can be written as
\begin{equation}\label{t}
p_i(s_k^\pm) = p_i^0 \pm t \, \sqrt {\lambda_k} \, v_{ik} \, ,
\end{equation}
with $s_k$ is the $k^{th}$ set of neutron FFs, $t$ adapted to make the desired $T^2 = \Delta \chi^2_{\rm global}$ which is the tolerance for the
required confidence interval (C.L.) and $t=T$ in the quadratic approximation.

Using the method we mentioned above, we accompany the construction of the QCD fit by reliable estimation of uncertainty.
Finally uncertainties of any observables ${\cal O}$, which can be the neutron FFs or reduced cross sections in our case, in the Hessian method can calculate as~\cite{Martin:2002aw,Martin:2009iq,Pumplin:2001ct}
\begin{equation}\label{Delta-F1}
\Delta {\cal O} = \frac{1}{2} \left[\sum_{k=1}^n({\cal O}(s_k^+) - {\cal O}(s_k^-))^2\right]^{\frac{1}{2}} \,.
\end{equation}
In above equation, ${\cal O}(s_k^+)$ and ${\cal O}(s_k^-)$ are the value of ${\cal O}$ extracted from the input set of parameters $p_i(s_k^\pm)$ obtained from Eq.~\eqref{t}.
In this paper, we follow the standard Hessian method to calculate the neutron FFs error band as well as the corresponding observables such as the reduced cross sections.
The evolved neutron FFs are attributive functions of the input parameters obtained in the QCD fit procedure at the scale $Q_0^2$, then their
uncertainty can be written applying the standard Hessian method 
\begin{equation}\label{Delta-F}
\Delta {\cal O} = \left[\Delta \chi^2_{\rm global} \, \sum_{i, j=1}^k \frac{\partial {\cal O}}{\partial p_i} \, C_{ij} \, \frac{\partial {\cal O}}{\partial p_j}\right]^{\frac{1}{2}} \,.
\end{equation}
The $\Delta \chi^2$ values determine the confidence region, and it is calculated so that the confidence level (C.L.) $P$ becomes the one-$\sigma$-error range
($P = 0.68$) for a given number of parameters ($p_{i=N}$) by assuming the normal distribution in the multi-parameter space.
Since the neutron FFs are provided with many parameters, so that the $\Delta \chi^2_{\rm global}$ value should be calculated.

Assuming correspondence between the confidence level (C.L.) of a normal distribution in multi-parameter space and the one of a $\chi^2$ distribution with $N$ degree of freedom, one can define the probability density function as

\begin{equation}
P_N(\chi^2) = \frac{ (\chi^2)^{\frac{N}{2} - 1}}{ 2^{\frac{N}{2}} \Gamma (\frac{N}{2}) }  e^{\frac{-\chi^2}{2}}  ~,
\end{equation}

then the confidence level $P$ can be obtain as

\begin{equation}\label{P6895}
\int_{0}^{\Delta \chi^2 } P_N(\chi^2) d \chi^2 = P (\approx 0.68)~,
\end{equation}
and similarly for the 90th percentile we have  $P = 0.90$.
The parameter number in our analysis is eight ($N = 8$), and it leads to $\Delta \chi^2$ = 9.27.
The uncertainty of a neutron FFs with respect to the optimized parameters $p_{i=N}$ is then calculated using Eq.~\eqref{Delta-F} by using Hessian matrices and assuming mentioned linear error propagation. For the neutron FFs uncertainty estimation, one can analytically calculate the gradient terms in Eq.~\eqref{Delta-F} at the initial scale $Q_0^2$ = 1 GeV$^2$. For the estimation at arbitrary Q$^2$, each gradient term is evolved by the DGLAP evolution kernel, and then the neutron FFs uncertainties as well as the uncertainties for any other observables such as cross sections are calculated. Here we calculate the neutron FFs uncertainty with $\Delta\chi^2 = 1$ and $9.27$ which is the most appropriate choice. 
The Hessian method discussed in the present analysis has been used for estimating {\tt TKAA16} NNLO polarized PDFs~\cite{Shahri:2016uzl} as well as {\tt KA15} nuclear PDFs analysis~\cite{Khanpour:2016pph}. The details of the uncertainty analysis are discussed in details in Refs.~\cite{Martin:2002aw,Martin:2009iq,Pumplin:2001ct}.

\begin{figure*}[htb]
	\vspace*{0.5cm}
	\includegraphics[clip,width=0.450\textwidth]{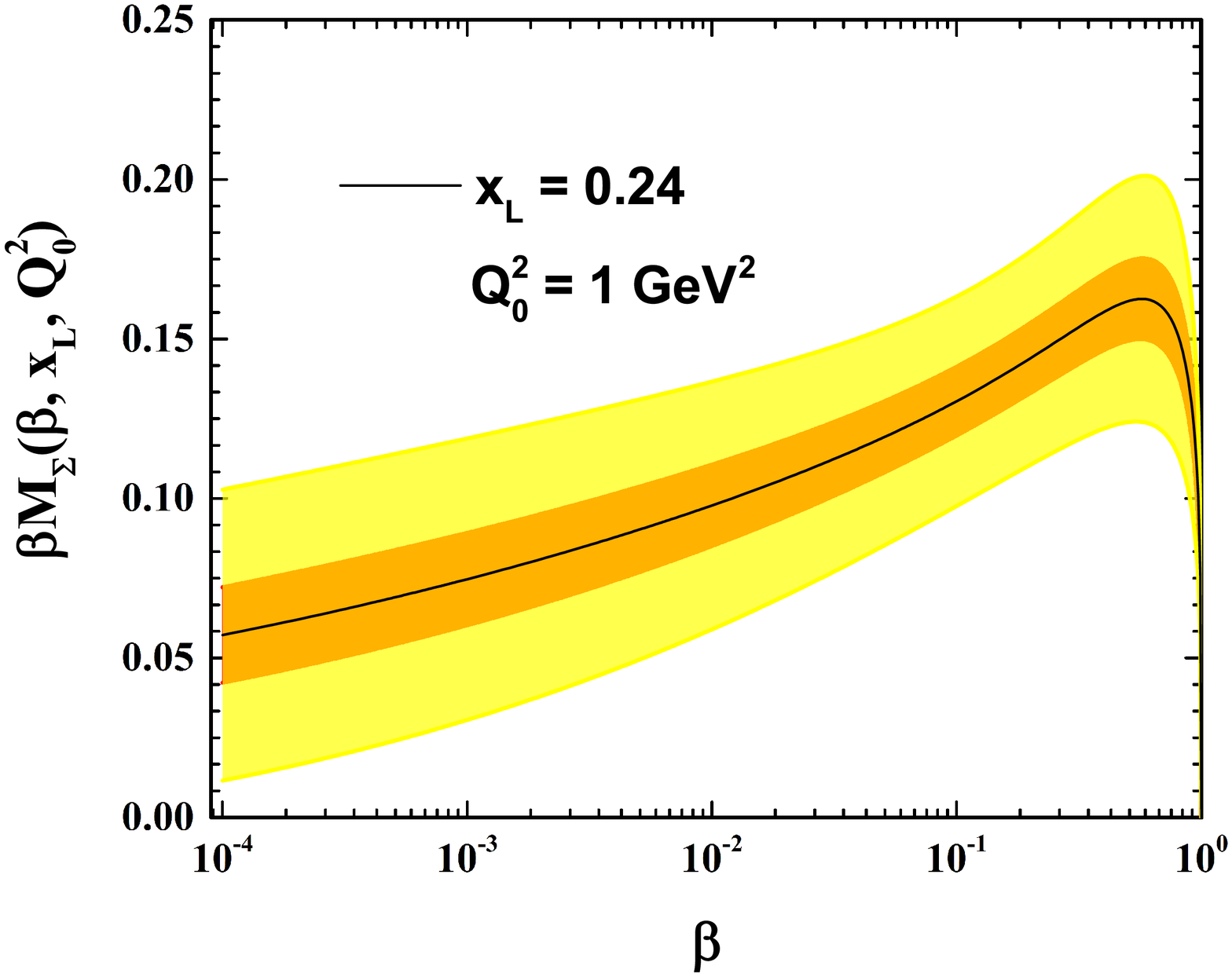}
	\includegraphics[clip,width=0.450\textwidth]{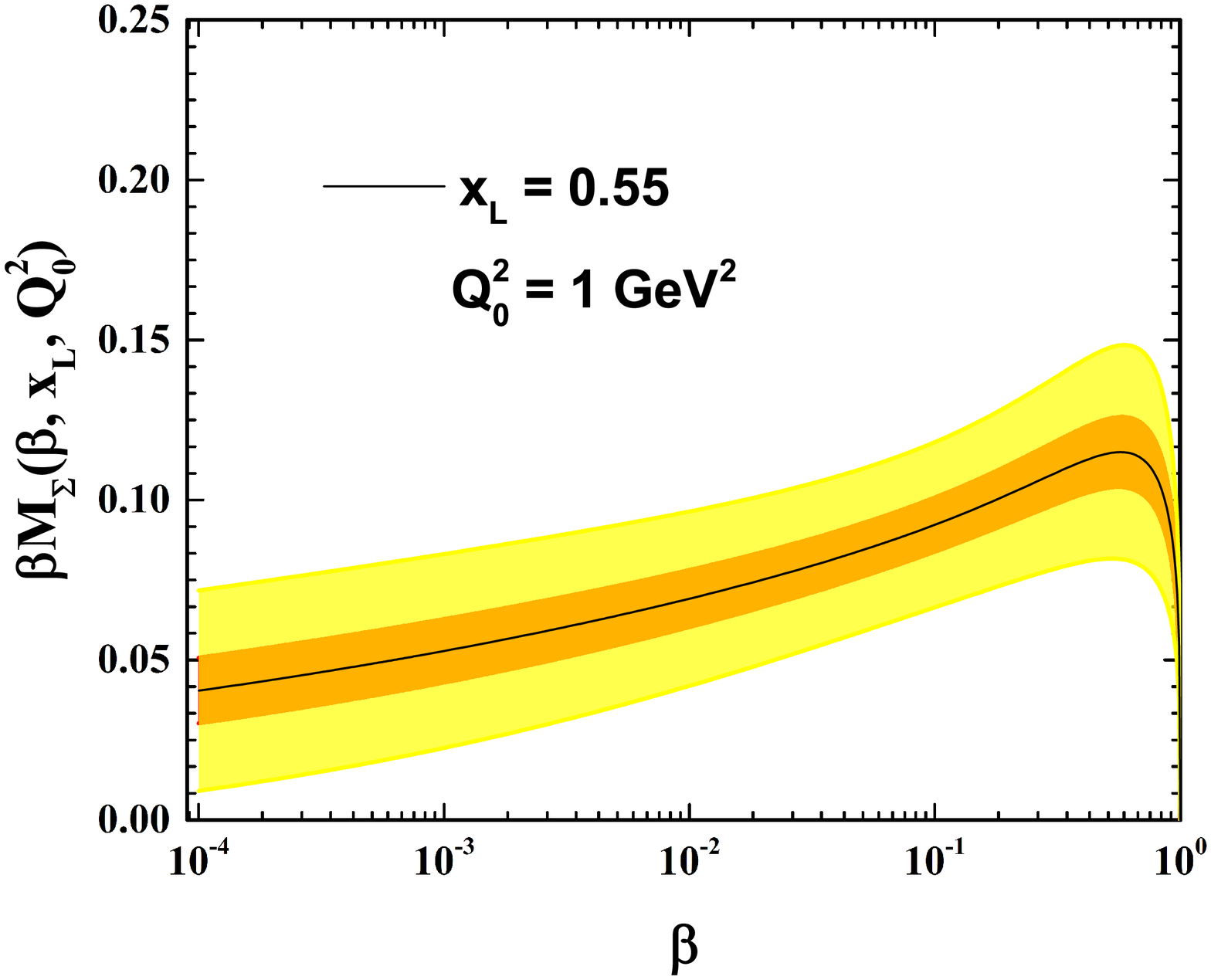}
	\includegraphics[clip,width=0.450\textwidth]{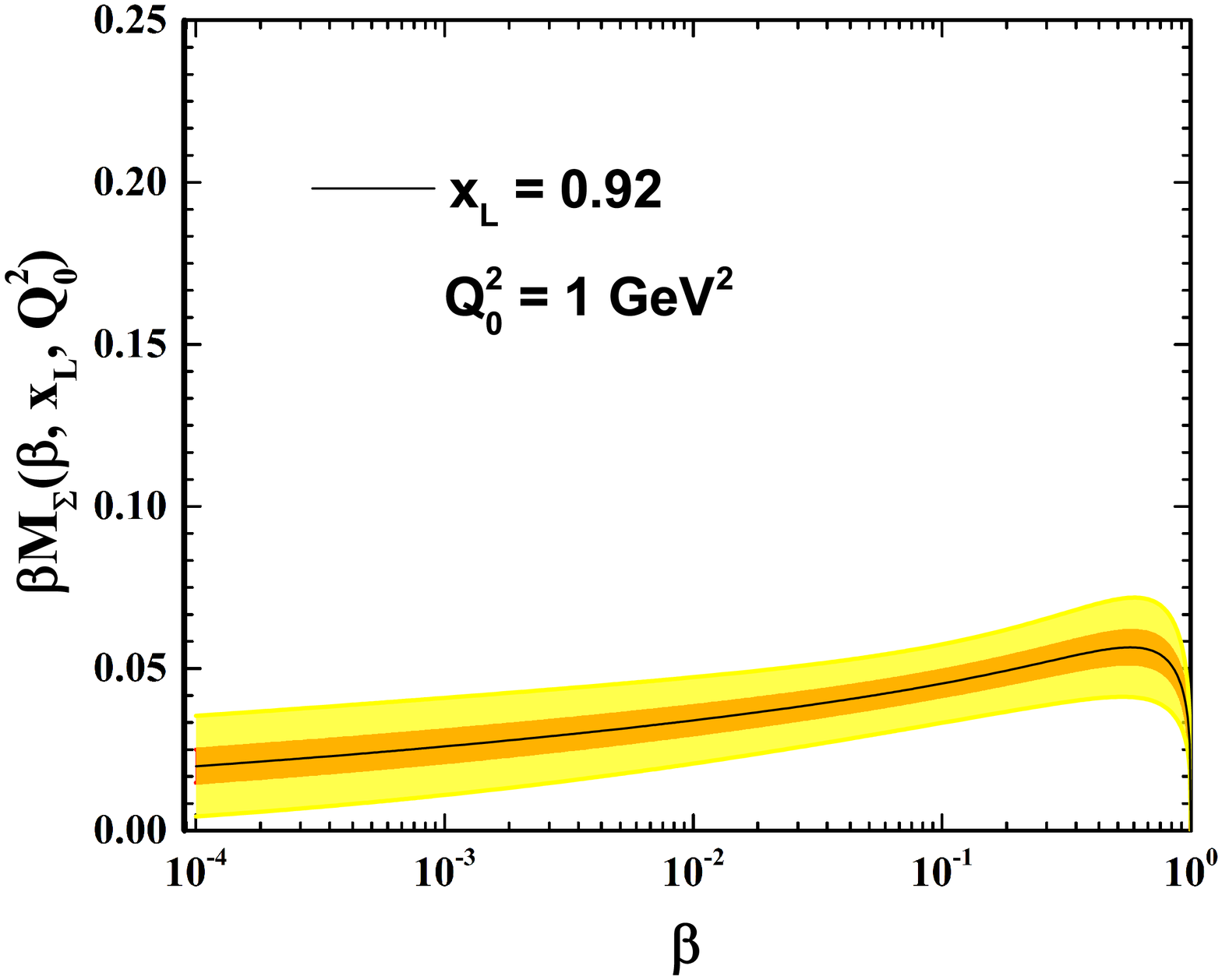}
	\begin{center}
		\caption{\small (Color online) The singlet momentum distribution as a function of $\beta$ at the input scale Q$_0^2$ = 1 GeV$^2$ and for three representative bins of $x_L$ = 0.24, 0.55 and 0.92. The error bands are obtained with the Hessian methods (see the text).
			\label{Fig-M-xL}}
	\end{center}
\end{figure*}
\begin{figure*}[htb]
	\vspace*{0.5cm}
	\includegraphics[clip,width=0.450\textwidth]{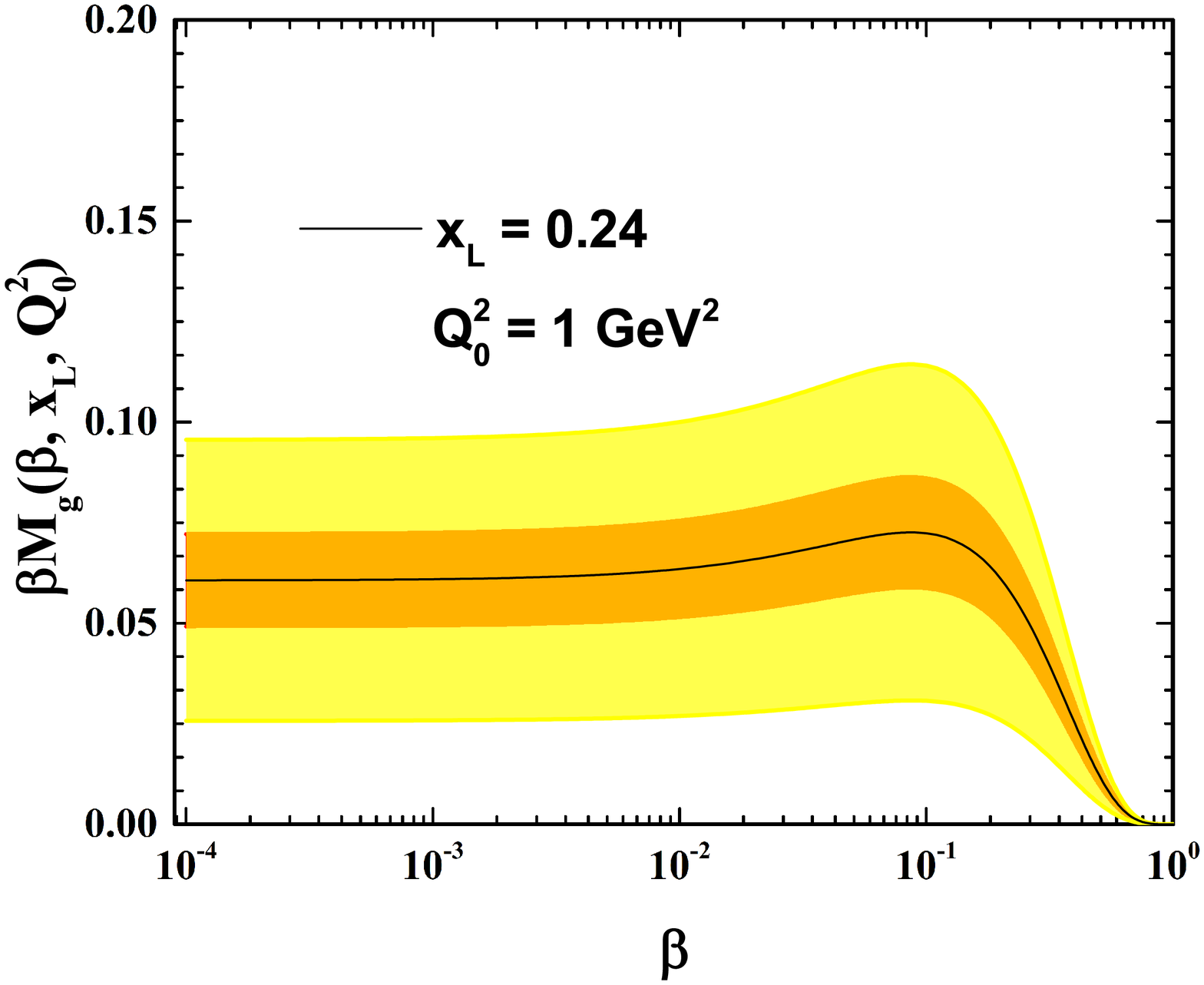}
	\includegraphics[clip,width=0.450\textwidth]{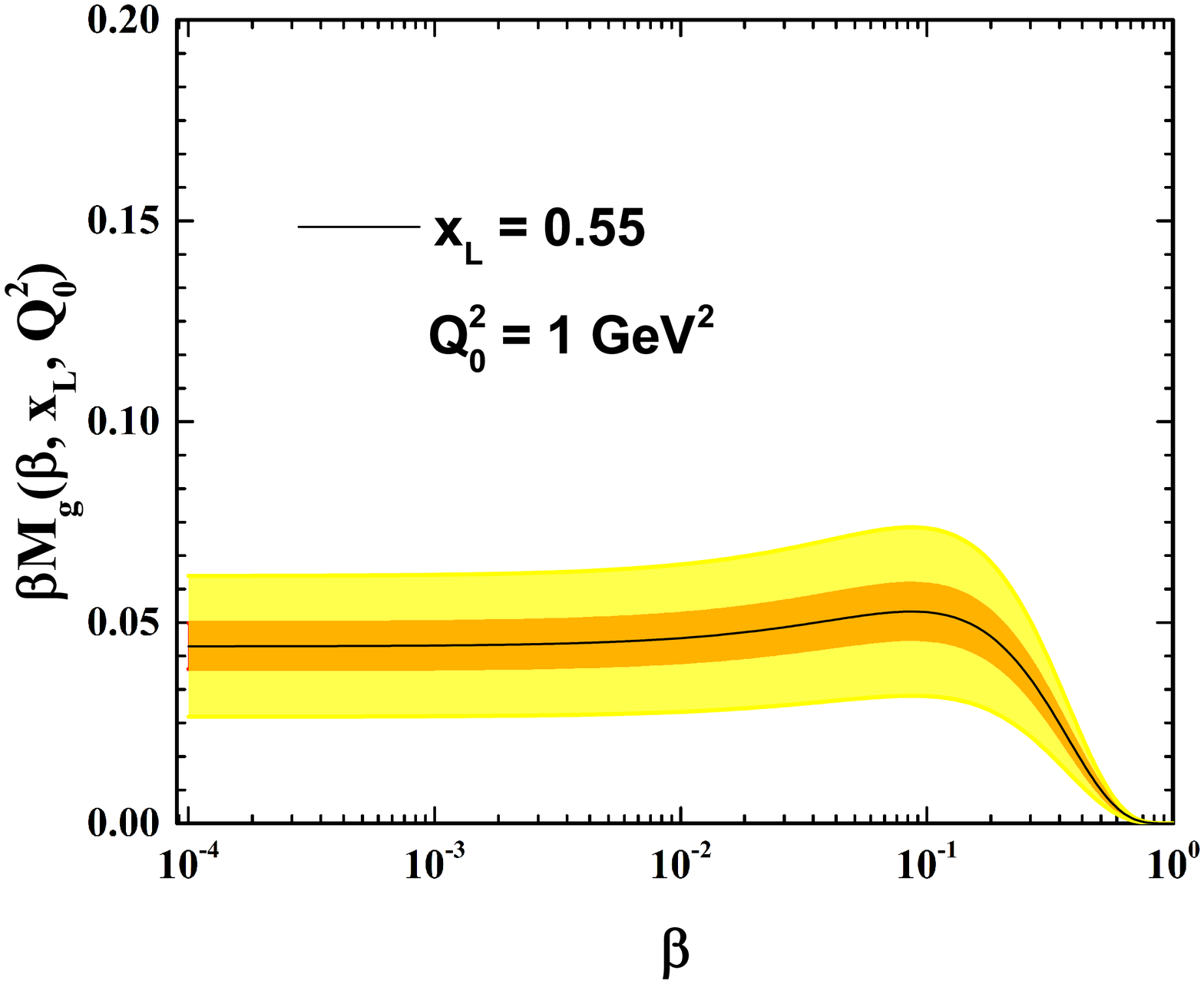}
	\includegraphics[clip,width=0.450\textwidth]{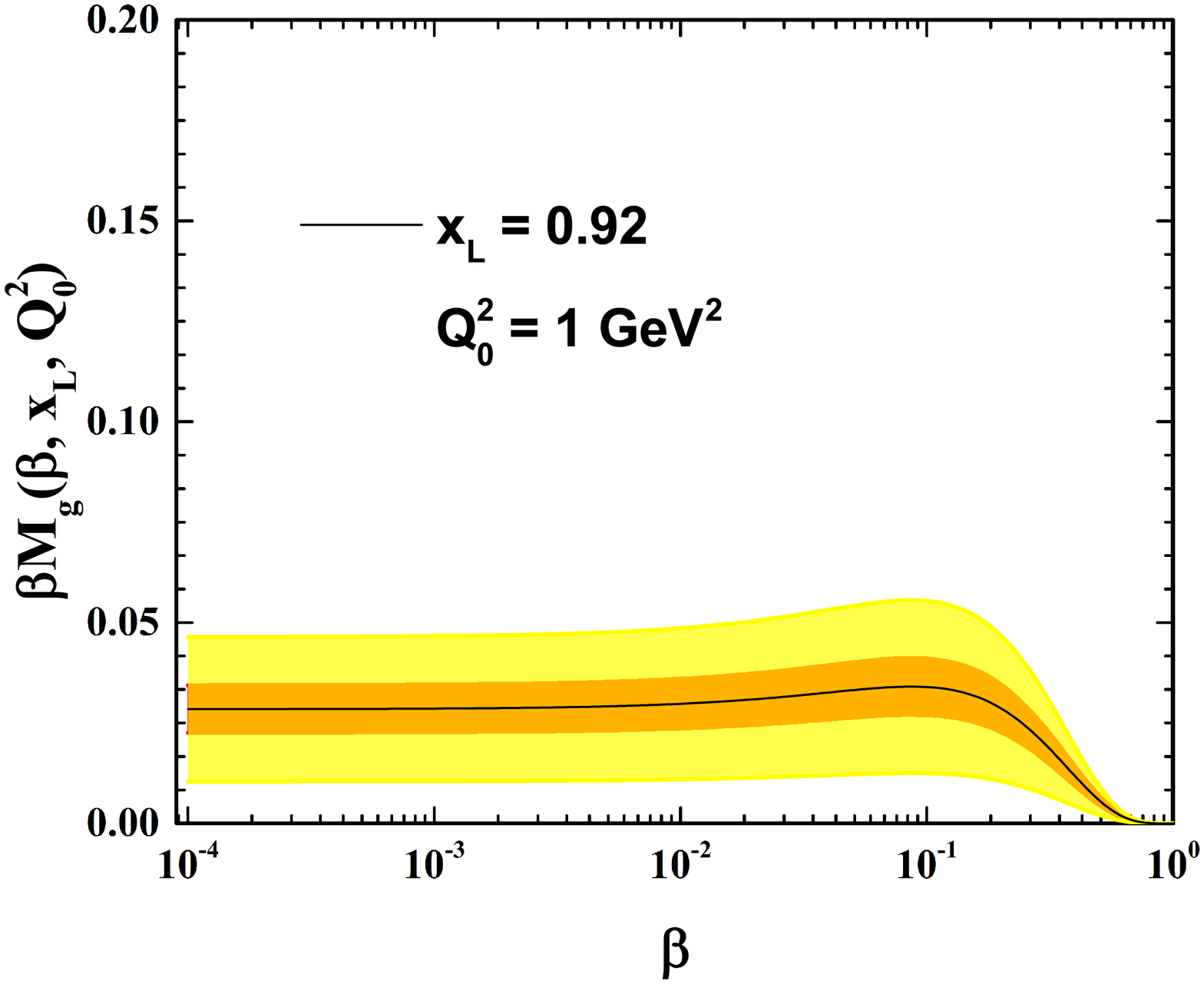}
	\begin{center}
		\caption{\small (Color online) The gluon momentum distribution as a function of $\beta$ at the input scale Q$_0^2$ = 1 GeV$^2$ and for three representative bins of $x_L$ = 0.24, 0.55 and 0.92. The error bands are obtained with the Hessian methods (see the text).
			\label{Fig-g-xL}}
	\end{center}
\end{figure*}

%
\section{Results and discussions}\label{sec:Results}

We are in a position to describe the details and all techniques we used for the parametrizations of neutron FFs in {\tt SKTJ17} global analysis. 
The minimization is carried out with respect to the set of parameters in Eq.~\eqref{eq:PDFQ0}, $\{p_i\} = \{a_i, b_i, c_i, {\cal N}_i, A_i, B_i, C_i, D_i \}$. The neutron FFs are evolved to the scales Q$^2$ > Q$_0^2$ relevant in experiment.
Like for the case of PDFs parameterization, particular functional form and the value for Q$_0^2$ are not too crucial.
The parameterization at the input scale should be flexible enough to accommodate all DIS data within their ranges of uncertainties.
As we mentioned, our input distributions in Eq.~\eqref{eq:PDFQ0} follow the standard form used in fits to DIS data.
In addition to our much more flexible input parametrization presented in Eq.~\eqref{eq:PDFQ0},
we have repeated our QCD fit with alternative parametrizations, some of them even more flexible than the one we choose.
For example, we have also included $\sqrt{x}$ terms, both for the singlet and gluon distributions, even allowing the fit to vary
them. We have found no significant improvement in the quality of the fit to data or changes of the uncertainty bands.
This indicates that the present H1 and ZEUS leading neutron production data is not really able to discriminate between
various forms of the input distributions.

As will be seen from our results presented in this section, we found that our input distributions in Eq.~\eqref{eq:PDFQ0} could be considered as good parametrizations to the leading neutron production experimental data. 

The parameter values $\{p_i\}$ of the next-to-leading order input neutron FFs at Q$_0^2$ = 1 GeV$^2$ obtained from the best fit to the combined H1 and ZEUS leading neutron data sets are presented in Table~\ref{fit-parameters}.

\begin{table*}[htbp]
	\caption{ Parameter values $\{p_i\}$ for {\tt SKTJ17} QCD analysis at the input scale Q$_0^2$ = 1 GeV$^2$ from the combined H1 and ZEUS data sets. The values without errors
		have been fixed after the first minimization since the data do not constrain these unknown parameters well enough. The details of the $\chi^2$ analysis and
		the constraints applied to control the  neutron FFs parameters are contained in the text. \label{fit-parameters}}
	\begin{tabular}{l|cc|ccccccc}
		\hline  \hline
		Parameters & $\beta M^N_{\Sigma/P} (\beta, Q_0^2, x_L)$  & $p_i \pm \delta p_i$  &   $\beta M^N_{g/P} (\beta, Q_0^2, x_L)$  & $p_i \pm \delta p_i$ \\   \hline  \hline
		$a$  & $a_q$ & $0.116 \pm 0.031$ & $a_g$ &  $0.0^*$    \\
		$b$  & $b_q$ & $0.260^*$ & $b_g$ &  $4.884^*$    \\
		$c$  & $c_q$ & $0.523^*$ & $c_g$ &  $9.969^*$    \\
		${\cal N}$  & ${\cal N}_q$ & $0.245 \pm 0.023$ &${\cal N}_g$ &  $0.130 \pm 0.027$    \\
		$A$  & $A_q$ & $0.0^*$ & $A_g$ &  $0.201^*$    \\
		$B$  & $B_q$ & $1.430 \pm 0.092 $ & $B_g$ &  $1.740 \pm 0.117$    \\
		$C$  & $C_q$ & $12.071 \pm 2.270 $ & $C_g$ &  $29.865^*$    \\
		$D$  & $D_q$ & $5.307 \pm 0.390 $ & $D_g$ &  $6.733 \pm 0.646$    \\  \hline  \hline
	\end{tabular}
\end{table*}

Parameters marked with ($^*$) are fixed. This is due to that these parameters are only very weakly determined by the fit, consequently we fixed them to their preferred values. For the sea quark density we set $A_q$ to $0$ and for the gluon density we set $a_g$ to $0$ in Eq.~\eqref{eq:PDFQ0}. These only marginally limit the freedom in the functional form. We found that singlet small-$x_L$ coefficient $A_q$ as well as gluon small-$\beta$ coefficient $a_g$ is determined with rather large error and also compatible with zero, so that we fixed them to these values. These are because there are no enough data sensitive to smaller values of $\beta$ and $x_L$. Moreover 
we found that the factor $( 1 + c_i \, \beta )$ in {\tt SKTJ17} parametrization provides flexibility to obtain a good description of the data. Thus, we will make
use of the $c_i$ coefficients for the sea quark and gluon densities. The parameters $B_q$ and $D_q$ always came out close to $B_g$ and $D_g$, so one can set them equal. In order to let  enough flexibility to the sea quark and gluon densities, we prefer them to vary differently in the QCD fit.  
In total this leaves us with 8 free parameters in the {\tt SKTJ17} QCD fit, (5 for sea quarks and 3 for the gluon density), which we include later on also in our uncertainty estimates. We also tried to relax the imposed constraints discussed above, but found that present leading neutron data are not really sensitive to them.
We find ${\chi}^{2}/{\rm {d.o.f.}} = 349.16 / 495 = 0.705$ which yields an acceptable fit to the experimental data.

%
\subsection{{\tt SKTJ17} neutron FFs and their uncertainties}\label{sec:SKTJ17nFFs}

Our newly obtained singlet and gluon momentum distributions at the input scale Q$_0^2$ = 1 GeV$^2$ are shown in Figs.~\ref{Fig-M-xL} and \ref{Fig-g-xL} along with estimates of their uncertainties using the
Hessian methods for a tolerance of $\Delta \chi^2 =1$ and $9.27$. The results presented for three representative bins of $x_L$ = 0.24, 0.55 and 0.92. 
The inner error band is obtained with the standard ``parameter-fitting'' criterion, by the choice of tolerance  T = $\Delta \chi^2 =1$ for the 68\% (one-sigma) confidence
level (C.L.) limit while the outer one is obtained with the choice of tolerance T = $\Delta \chi^2 = 9.27$ using Eq.~\ref{P6895}. 
The main conclusion that can be drawn about the gluon and singlet distributions from {\tt SKTJ17} analysis is that the distributions are important
at large $\beta$. As was stated earlier in Sec.\ref{sec:Results}, their behavior cannot be precisely determined yet from the available leading neutron production data.
In particular, the behavior of the exponent of the $(1 - \beta)$ factors in the parametrization, $b_q$ and $b_g$.

\begin{figure*}[htb]
	\vspace*{0.5cm}
	\includegraphics[clip,width=0.50\textwidth]{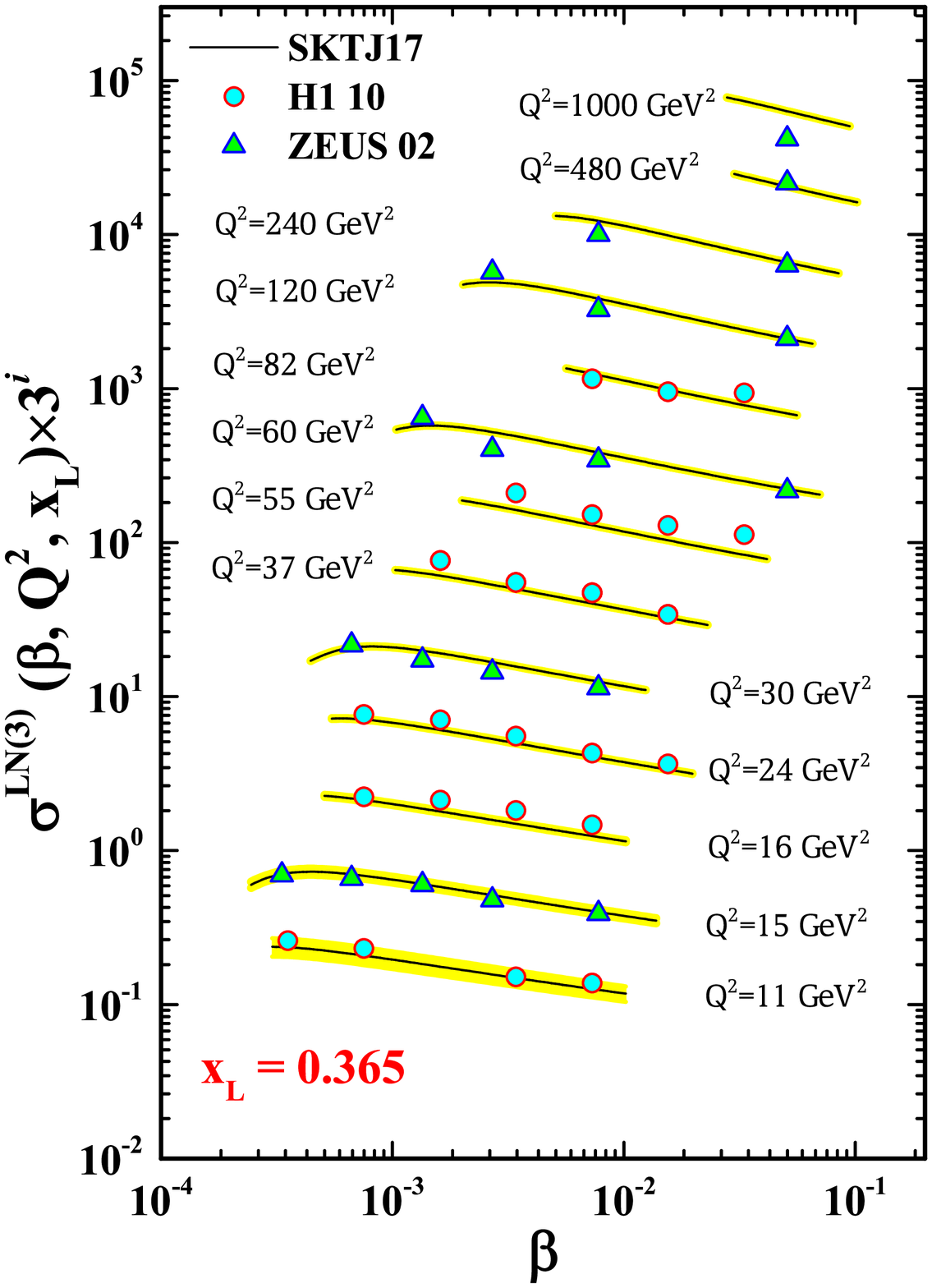}
	\includegraphics[clip,width=0.50\textwidth]{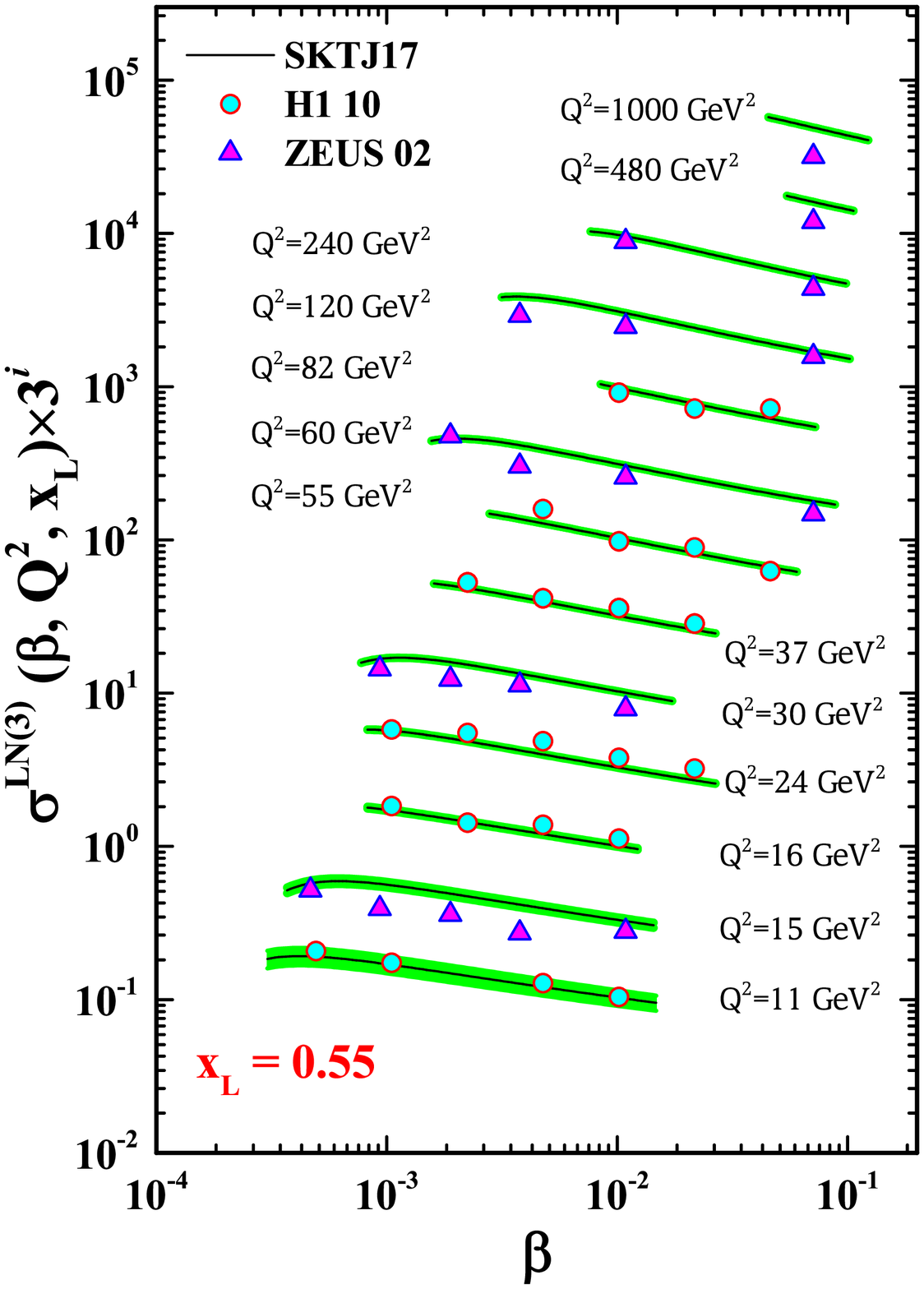}\vspace{-2cm}
	\includegraphics[clip,width=0.50\textwidth]{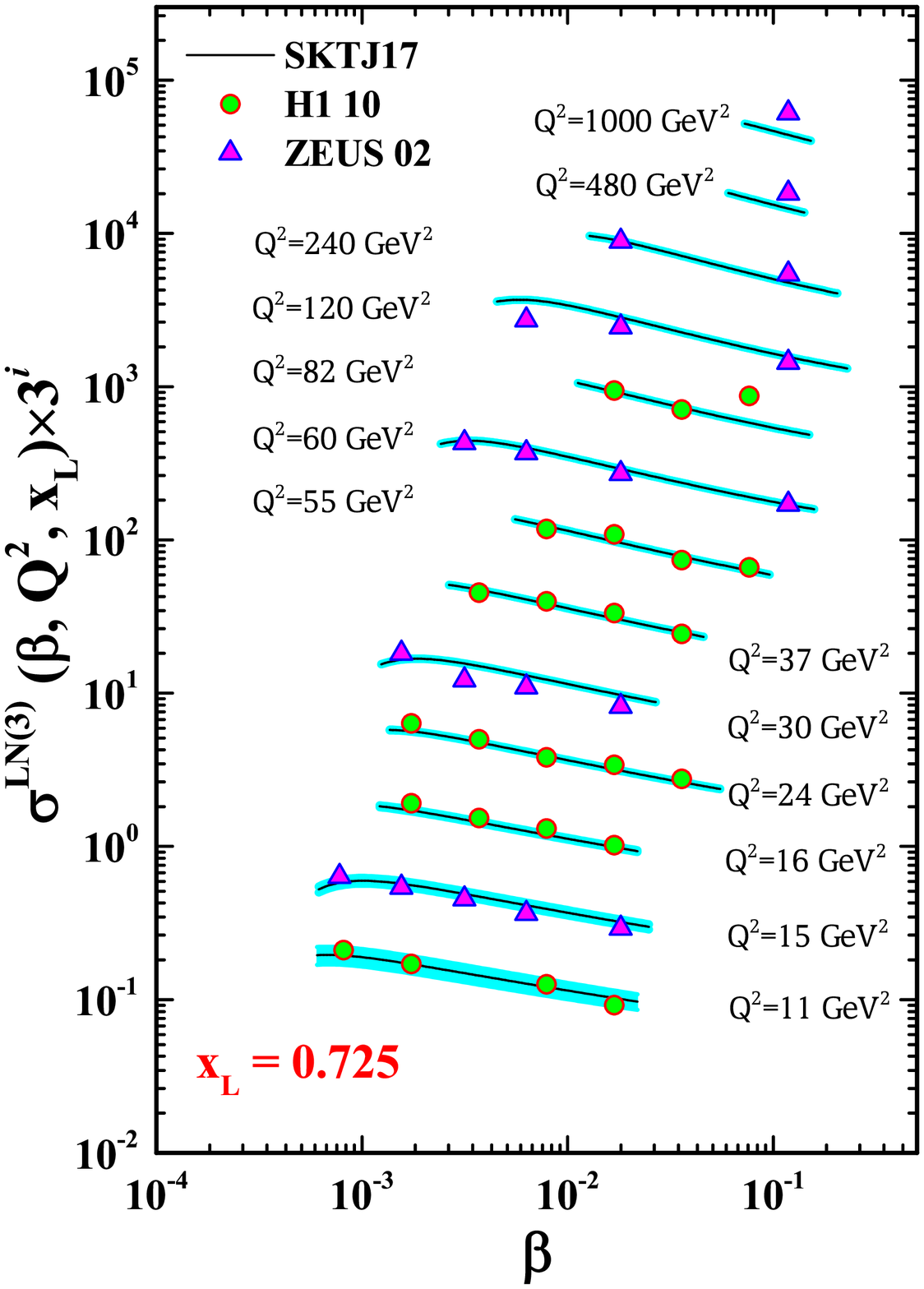}
	\vspace{-1cm}
	\begin{center}
		\caption{\small (Color online) The reduced cross sections $\sigma_r^{\rm LN(3)} (\beta, x_L, Q^2)$ as a function of $\beta$ for some selected values of Q$^2$ (in GeV$^2$ units) and for three representative bins of $x_L$ = 0.365, 0.550 and 0.725.
			To facilitate the graphical presentation we have plotted $\sigma_r^{\rm LN(3)} (\beta, x_L, Q^2) \times 3^i$.  \label{Sigma-beta}}
	\end{center}
\end{figure*}

\begin{figure*}[htb]
	\vspace*{0.5cm}
	\includegraphics[clip,width=0.50\textwidth]{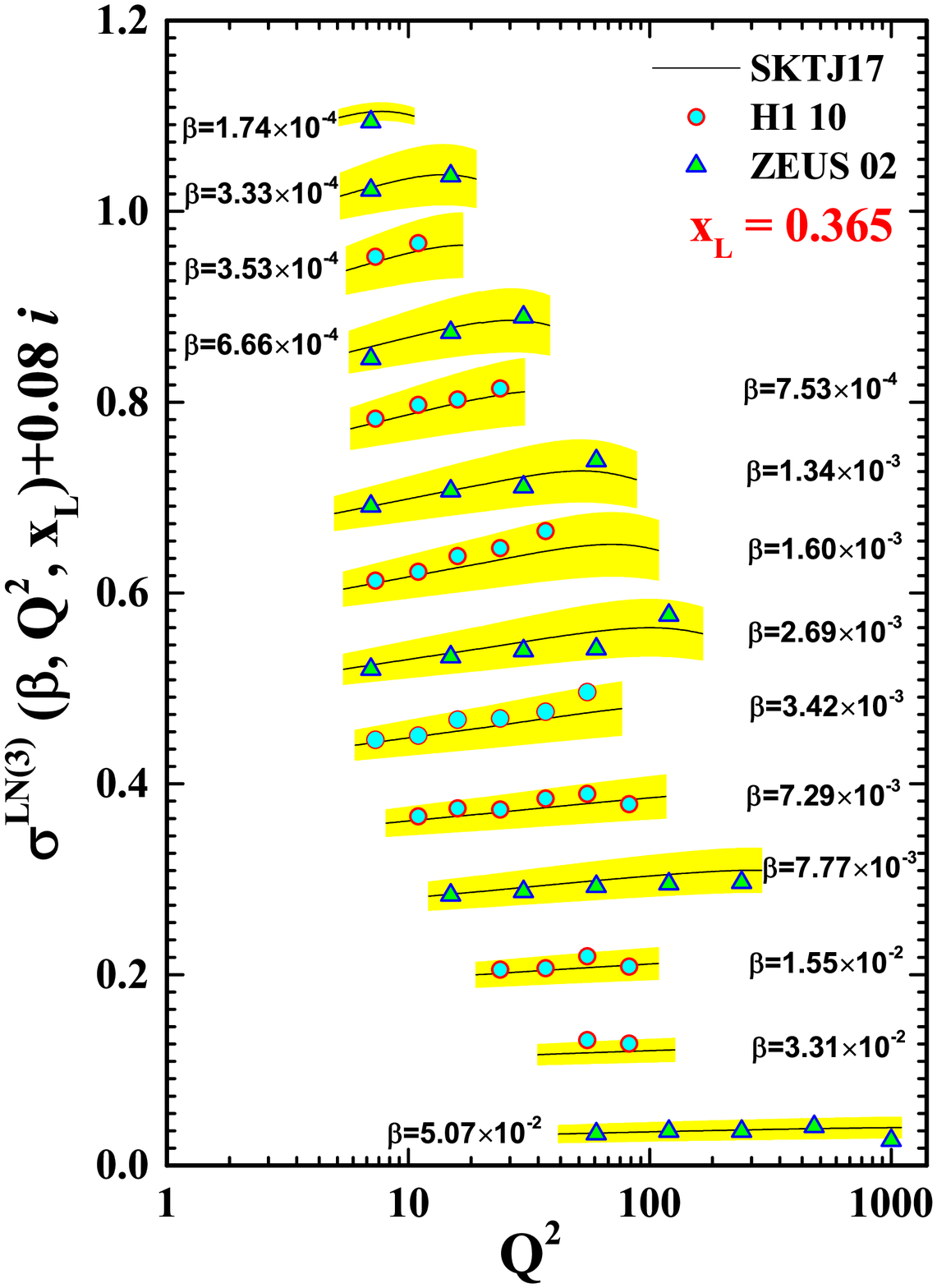}
	\includegraphics[clip,width=0.50\textwidth]{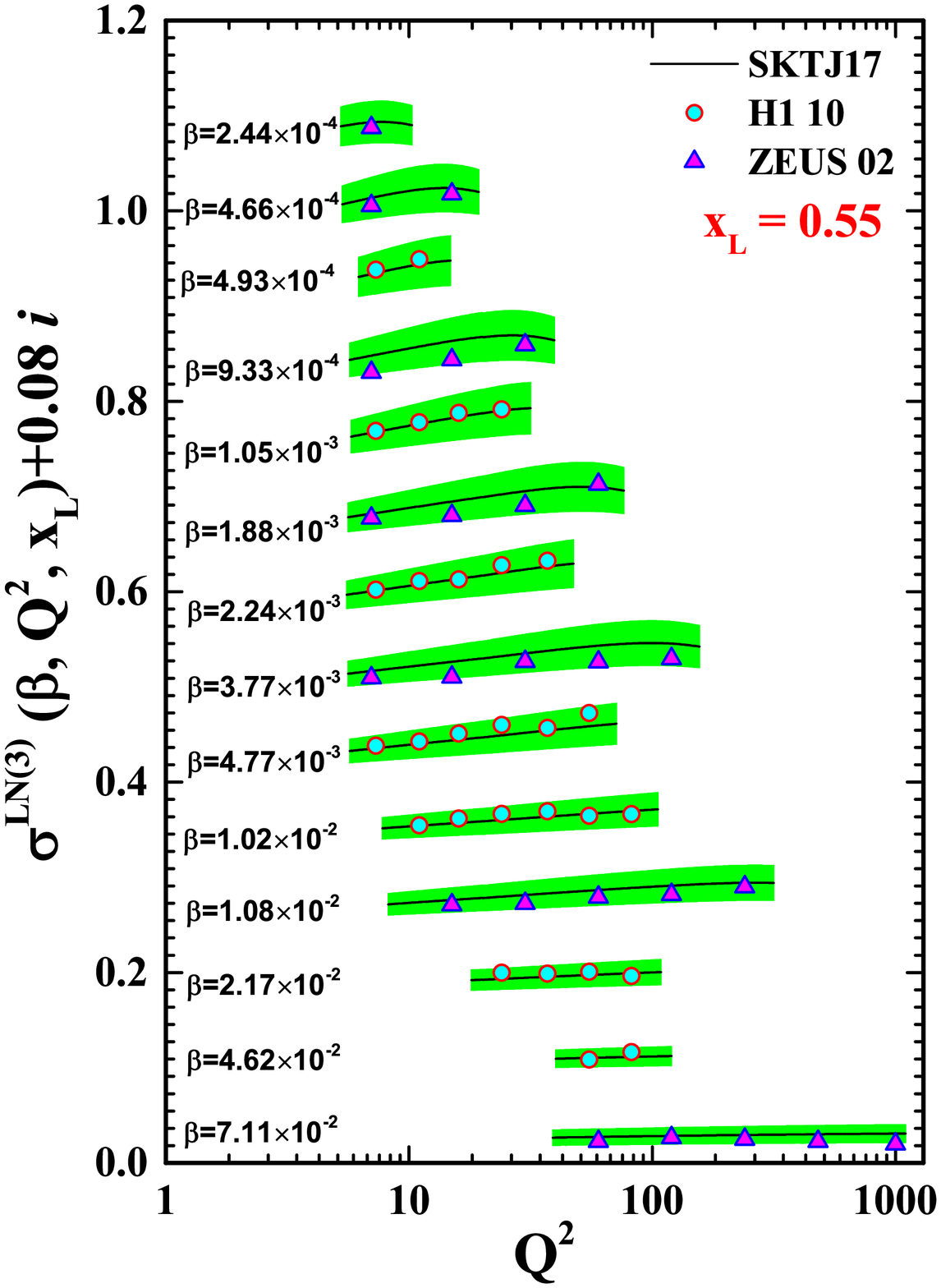}\vspace{-2cm}
	\includegraphics[clip,width=0.50\textwidth]{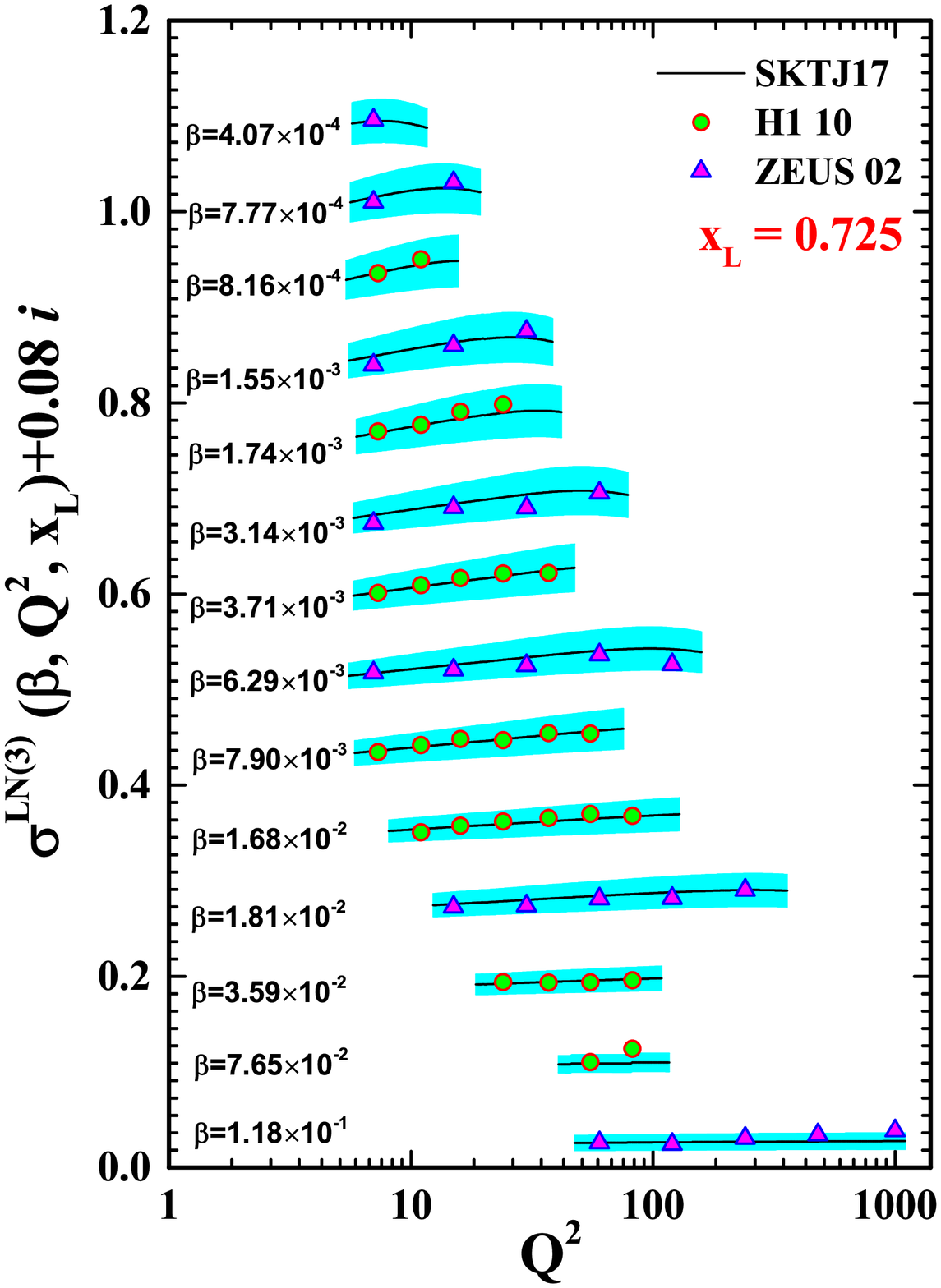}
	\vspace{-1cm}
	\begin{center}
		\caption{\small (Color online) The reduced cross section $\sigma_r^{\rm LN(3)} (\beta, x_L, Q^2)$ as a function of Q$^2$ for some selected values of $\beta$ and for three representative bins of $x_L$ = 0.365, 0.550 and 0.725. To facilitate the graphical presentation we have plotted $\sigma_r^{\rm LN(3)} (\beta, x_L, Q^2) + 0.08 i$. \label{Sigma-Q}}
	\end{center}
\end{figure*}

%
\subsection{ Comparison to leading neutron data  }\label{sec:Comparisondata}

In order to check the reliability of the distributions obtained in our analysis, in the following we compare results obtained using our best parametrization in Eq.~\eqref{eq:PDFQ0} with the leading neutron production data sets presented by the H1 and ZEUS collaboration in which have been included in the {\tt SKTJ17} fit. 
In Figs.~\ref{Sigma-beta}, {\tt SKTJ17} theory predictions for the reduced cross section $\sigma_r^{\rm LN(3)} (\beta, Q^2, x_L)$ are plotted as a function of $\beta$ for some selected values of Q$^2$. For better description of the fit quality for different region of $x_L$, three representative bins of $x_L$ = 0.365, 0.55 and 0.725 are shown. The reduced cross section $\sigma_r^{\rm LN(3)} (\beta, Q^2, x_L)$ is scaled by a factor of $3^i$ for better visibility in the plots. 
In order to see the fit quality, the leading neutron production data from H1 and ZEUS collaborations~\cite{Aaron:2010ab,Chekanov:2002pf} also added to these plots. 
From the figures it is clear that {\tt SKTJ17} QCD fit based on hard-scattering formula in Eq.~\eqref{eq:reduced} together with the neutron FFs initial
conditions in Eq.~\eqref{eq:PDFQ0} are in acceptable agreement with the H1 and ZEUS data. The plots also show that our results describe the data well, down to the lowest accessible value
of Q$^2$ as well as for different region of $x_L$.

In order to study the scale dependence of H1 and ZEUS leading neutron data, we have plotted the reduced cross sections $\sigma_r^{\rm LN(3)} (\beta, Q^2, x_L)$ as a function of Q$^2$  in Fig.~\ref{Sigma-Q} for some selected values of $\beta$ and for three representative bins of $x_L$ = 0.365, 0.550 and 0.725. The reduced cross section $\sigma_r^{\rm LN(3)} (\beta, Q^2, x_L)$ is scaled by a factor of $0.08 i$ for better visibility in the plots. One can conclude our results show that the scale dependence induced by the evolution equations of Eq.~\eqref{eq:DGLAP} is perfectly consistent with the leading neutron production data. The results clearly show that one can use the fracture functions approach to describe semi-inclusive hard processes in perturbative QCD at the kinematic region covered by electron-proton collider HERA and hadron colliders.  

\begin{figure*}[htb]
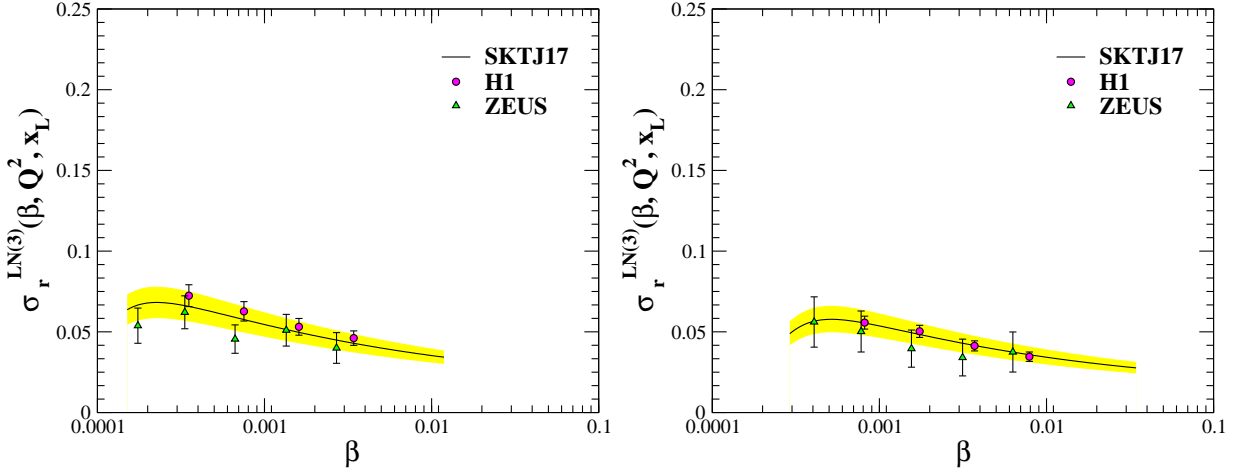

	\vspace*{0.5cm}
	\includegraphics[clip,width=0.45\textwidth]{4plot-ave-sb1.eps}
	\includegraphics[clip,width=0.45\textwidth]{4plot-ave-sb2.eps}
	\begin{center}
		\caption{\small (Color online) {\tt SKTJ17} theory predictions for the reduced cross sections $\sigma_r^{\rm LN(3)} (\beta, Q^2, x_L)$ as a function of $\beta$.
		The H1 (ZEUS) data correspond to Q$^2$ = 7.3 (7.0) GeV$^2$, and $x_L$= 0.365 (0.370) in the left panel and $x_L$= 0.725 (0.730) in the right panel. The error bars associated with the H1 and ZEUS data points include systematic and statistical uncertainties, being the total experimental error evaluated in quadrature. \label{4plot-ave-sb}}
	\end{center}
\end{figure*}

In Fig.~\ref{4plot-ave-sb}, our theory predictions for the reduced cross sections $\sigma_r^{\rm LN(3)} (\beta, Q^2, x_L)$ shown as a function of $\beta$. The H1 (ZEUS) data correspond to Q$^2$ = 7.3 (7.0) GeV$^2$, and $x_L$= 0.365 (0.370) in the left panel and $x_L$= 0.725 (0.730) in the right panel. As shown in the plots, we obtain remarkable agreement with the data in the common $x_L$ and $\beta$ range. The plots also clearly show that our approach based on the fracture functions formalism allow a unified description of leading neutron deep inelastic cross sections.

\begin{figure*}[htb]
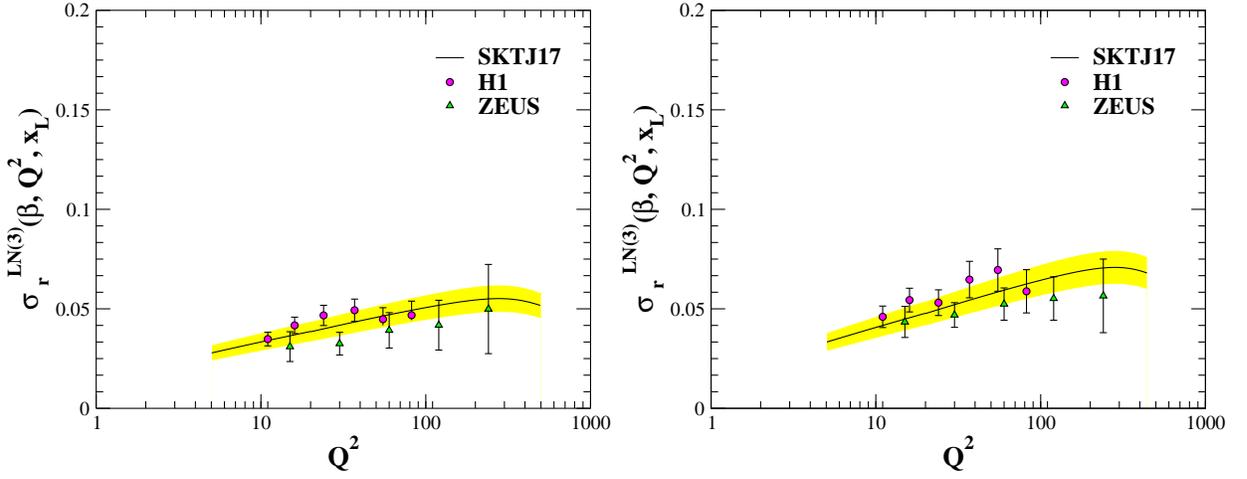

	\vspace*{0.5cm}
	\includegraphics[clip,width=0.45\textwidth]{4plot-ave-sq1.eps}
	\includegraphics[clip,width=0.45\textwidth]{4plot-ave-sq2.eps}
	\begin{center}
		\caption{\small (Color online) {\tt SKTJ17} theory predictions as a function of Q$^2$ (in GeV$^2$ units). 
		The error bars associated with the H1 and ZEUS data points include systematic and statistical uncertainties, being the total experimental error evaluated in quadrature. In the right panel, the H1 (ZEUS) data correspond to $\beta = 7.29 \, (7.77) \times 10^{-3}$ and $x_L = 0.365 \, (0.370)$. The H1 (ZEUS) data in the left panel correspond to $\beta = 1.02 \, (1.08) \times 10^{-2}$ and $x_L = 0.545 \, (0.550)$. \label{4plot-ave-sq}}
	\end{center}
\end{figure*}

For completeness, we finally show {\tt SKTJ17} theory predictions as a function of Q$^2$ for the reduced cross sections $\sigma_r^{\rm LN(3)} (\beta, Q^2, x_L)$ with a representative selection of H1 and ZEUS data in Fig.~\ref{4plot-ave-sq}. In the right panel, the H1 (ZEUS) data corresponds to $\beta = 7.29 \, (7.77) \times 10^{-3}$ and $x_L = 0.365 \, (0.370)$. The H1 (ZEUS) data in the left panel corresponds to $\beta = 1.02 \, (1.08) \times 10^{-2}$ and $x_L = 0.545 \, (0.550)$.
The results demonstrate that {\tt SKTJ17} theory predictions can provide good description of the HERA leading neutron spectra at all kinematics.

In this section, we turned to present our perturbative predictions for the reduced cross section and detail comparison with the available leading neutron production data.
Summarizing, our analysis provided a good description of H1 and ZEUS data for leading neutron production in DIS, as a function of $\beta$, $Q^2$ and $x_L$.
The analysis results presented in this section enabled us to established the models and parameters which are best able to well describe the existing leading neutron production data from H1 and ZEUS collaborations. In spite of the fact that excellent descriptions of the H1 and ZEUS leading neutron spectra are obtained over the entire range of $\beta$, $x_L$ and Q$^2$ covered by the data, new data could enable further constraints on the extracted neutron FFs. The success of the {\tt SKTJ17} global analysis performed here, stands for an explicit check of the pQCD framework in the fracture functions approach for the description of the leading neutron production processes.

%
\section{ Leading-baryons production at the LHC }\label{sec:LHC}

Let us here conclude by listing some further possible developments of the present framework as well as experimental efforts.
One of the important goals in high energy particle physics is to understand the production of leading-baryons which have large fractional longitudinal momentum $x_L \geq 0.3$.
Recent measurements of leading proton and neutron spectra in electron-proton collisions by H1 and ZEUS collaborations at HERA have open a new window on this subject.  
Very recently, H1 collaboration at HERA has been measured for the first time the photoproduction cross section for exclusive $\rho^0$ production associated with a leading neutron~\cite{H1:2015bxa}. Since there is no hard scale presented in exclusive $\rho^0$ production, one can use a phenomenological approach such as Regge
theory or color dipole formalism, to describe these kind of reactions~\cite{Ducati:2016jdg,Lebiedowicz:2016ryp,Goncalves:2015mbf}. 

Nowadays, our understanding of the hadron structure as well as the QCD dynamics have advanced with the successful operation and precise data at HERA collider. In addition to the HERA collider, the next generation of high energy and high luminosity electron-proton colliders, such as Large Hadron Electron Collider (LHeC)~\cite{AbelleiraFernandez:2012ni,AbelleiraFernandez:2012ty,Acar:2015cva} as well as Future Circular Hadron-Electron Collider (FCC-he)~\cite{Acar:2016rde} which are proposed to build on the same site with LHC, could help to study the leading-baryon processes.

Another possibility is the use of the hadronic colliders. One of the important
issue which have strong implications in the forward physics at hadron colliders, is the understanding of the leading neutron processes. 
A very rich program at the Large Hadron Collider (LHC) is being pursued in forward physics with sufficient experimental information~\cite{N.Cartiglia:2015gve,Albrow:2006xt}. 
Finally, the upcoming experiment at Jefferson Lab (JLAB) plans to take data on the production of leading protons in the $e n \to e p X$ process~\cite{Goncalves:2016uhj,TDIS-JLAB,R.A.Montgomery:2017hab}.
With the help of more and precise upcoming experimental data on such processes, a new era for theoretical understanding of strong
interactions in the soft, non-perturbative regime will be open~\cite{Goncalves:2016uhj}.

%
\section{Summary and Conclusion}\label{sec:Summary}

In the recent years, several dedicated experiments at the electron-proton collider HERA have collected high-precision data on the spectrum of leading-baryons carrying a large fraction of the proton's energy. However the experimental information on leading-baryons production in lepton DIS, $ep \to e^{\prime} B X$, is still rather scarce. In addition to these experimental efforts, much successful phenomenology has been developed in understanding the mechanism of leading-baryon productions. 
The presence of a leading-baryon in the final state of lepton DIS provides valuable information on the relationship between the soft and hard aspects of the strong interaction.

In this work we have presented {\tt SKTJ17} NLO QCD analysis of neutron FFs using available and up-to-date data from forward neutron production at HERA~\cite{Aaron:2010ab,Chekanov:2002pf}.
It is shown that an approach based on the fracture functions formalism allows us phenomenologically parametrize the neutron FFs.
We also have shown that a standard simple parametric form for this function gives a very accurate description of the available leading neutron production data.
Finally, one can conclude that our obtained results based on the fracture function approach agree well with the scale dependence of the leading neutron production data.
Completing such a picture is crucial as hadron colliders enter an era of new-generation of experimental data capable of testing this formalism.
In order to asses the uncertainties in the resulting neutron FFs and the corresponding observables, associated with the uncertainties in the data, we have made an extensive use of the Hessian method.

A {\tt FORTRAN} package containing {\tt SKTJ17} neutron FFs parameterization as well as the corresponding error set can be obtained via Email from the authors upon request. This {\tt FORTRAN} package also includes an example program to illustrate the use of the routines.

%
\section*{Acknowledgments}

The authors are especially grateful Garry Levman, Katja Kruger, Stefan Schmitt, Mohsen Khakzad from H1 and ZEUS collaborations for many useful discussions and comments.
We are also thankful to Federico Ceccopieri, Fatemeh Arbabifar, Muhammad Goharipour and S. Atashbar Tehrani for numerous informative discussions. 
Hamzeh Khanpour is indebted the University of Science and Technology of Mazandaran and the School of Particles and Accelerators, Institute for Research in Fundamental Sciences (IPM), to support financially this project. Fatemeh Taghavi-Shahri and Kurosh Javidan also acknowledge Ferdowsi University of Mashhad.


%
%

%

\end{document}